\documentclass[12pt]{article}
\usepackage[latin9]{inputenc}
\usepackage{amsmath}
\usepackage{amsthm}
\usepackage{graphicx}
\usepackage{setspace}
\usepackage[authoryear]{natbib}
\makeatletter

\DeclareTextSymbolDefault{\textquotedbl}{T1}
\providecommand{\tabularnewline}{\\}

\theoremstyle{definition}
 \newtheorem{example}{\protect\examplename}
\theoremstyle{plain}
\newtheorem{prop}{\protect\propositionname}

\@ifundefined{date}{}{\date{}}
\usepackage{url}% not crucial - just used below for the URL 

\newcommand{\blind}{0}

\providecommand{\examplename}{Example}
\providecommand{\propositionname}{Proposition}

\makeatother

\providecommand{\examplename}{Example}
\providecommand{\propositionname}{Proposition}

\begin{document}
\global\long\def\spacingset{ \global\long\global\long\global\long\def\baselinestretch{%
}
\small\normalsize}
 \if0\blind {

\title{\textbf{A Scalable Gaussian Process for Large-Scale Periodic Data}}

\author{Yongxiang~Li, Yuting~Pu\\
 {\small Department of Industrial Engineering and Management,} \\ 
 {\small Shanghai Jiao Tong University, Shanghai, China.} \\
 Changming~Cheng\\
 {\small State Key Laboratory of Mechanical System and Vibration,} \\ 
 {\small Shanghai Jiao Tong University, Shanghai, China.} \\
 Qian Xiao\thanks{Contact: Qian Xiao, qian.xiao@uga.edu, 310 Herty Dr, Department of Statistics, University of Georgia, Athens, GA, USA, 30602.} \\
{\small Department of Statistics,} \\
{\small University of Georgia, Athens, GA, USA}\\
}

\maketitle
} \fi

\if1\blind { \bigskip{}
 \vphantom{} \bigskip{}
 \vphantom{} \bigskip{}

\begin{center}
\textbf{\LARGE{}A Scalable Gaussian Process for Large-Scale Periodic
Data}{\LARGE\par}
\par\end{center}

\begin{center}
 
\par\end{center}

\medskip{}
 } \fi 
\begin{abstract}
The periodic Gaussian process (PGP) has been increasingly used to
model periodic data due to its high accuracy. Yet, computing the likelihood
of PGP has a high computational complexity of $\mathcal{O}\left(n^{3}\right)$
($n$ is the data size), which hinders its wide application. To address
this issue, we propose a novel circulant PGP (CPGP) model for large-scale
periodic data collected at grids that are commonly seen in signal
processing applications. The proposed CPGP decomposes the log-likelihood
of PGP into the sum of two computationally scalable composite log-likelihoods,
which do not involve any approximations. Computing the likelihood
of CPGP requires only $\mathcal{O}\left(p^{2}\right)$ (or $\mathcal{O}\left(p\log p\right)$
in some special cases) time for grid observations, where the segment
length $p$ is independent of and much smaller than $n$. Simulations
and real case studies are presented to show the superiority of CPGP
over some state-of-the-art methods, especially for applications requiring
periodicity estimation. This new modeling technique can greatly advance
the applicability of PGP in many areas and allow the modeling of many
previously intractable problems.\footnote{Supplementary materials for this article are available online.}
\end{abstract}

\section{Introduction \label{sec:intro}}

Periodic data, such as speech signals \citep{nielsen2017fast}, electrocardiogram
(ECG) signals \citep{chandola2011gaussian}, and vibration signals
\citep{fan2018noise}, are commonly encountered in scientific research
and industrial applications. Such signals are often analyzed via linear
models, e.g., the maximum likelihood pitch estimation (MLPE) method
\citep{wise1976maximum} and the noise resistant correlation (NRC)
method \citep{li2021extended}, or nonlinear models, e.g., the nonlinear
least square (NLS) method \citep{quinn1991estimating,nielsen2017fast}.
These methods can provide desirable estimations and predictions in
many applications, but they become less accurate when handling signals
with a low signal-to-noise ratio (SNR) and may require very long signals
to suppress the masking effect of strong background noises \citep{fan2018noise,li2021extended}.
A key reason is that they do not appropriately model the circulant
within-period correlation of periodic data, i.e., the autocorrelation
between any two points in one period.

The periodic Gaussian process (PGP) has been increasingly used in
recent years, especially for modeling signals under strong noise environments.
It can well model the circulant within-period correlation and thus
significantly improve the performance \citep{zhang2005time,chandola2011gaussian}.
For example, \citet{chandola2011gaussian} proposed a PGP-based change
point detection for the online monitoring of periodic time series,
which has superior performance compared with many current methods.
\citet{durrande2016detecting} adapted the framework of PGP to detect
the periodicity in the Arabidopsis genome, and \citet{guerin2020robust}
developed a robust object detection based on PGP filtering. \citet{koulali2021modelling}
proposed to use PGP for modeling geodetic time series to estimate
the secular velocity of selected GPS sites. PGP has also been used
for long-term forecasting of periodic processes and periodic error
control \citep{hajighassemi2014analytic,klenske2015gaussian}.

Despite its advantage in accuracy, PGP faces two challenges that hinder
its wide application. First, computing the likelihood of PGP requires
a high complexity of $\mathcal{O}\left(n^{3}\right)$, where $n$
is the signal length. It can be prohibitively slow when dealing with
moderate or large-scale data, e.g., spending several hours to process
10,000 data points on a normal computer. Second, PGP often requires
a known period or treats it as a tuning parameter. Yet, in real applications,
the true period is often unknown, and periodicity detection is often
of key interest. The current literature lacks efficient methods for
estimating periods in PGP, which is a challenging optimization problem
involving numerous local optimums.

To address the challenge on computation, various approximation methods
in the current GP literature may be adopted, which can be summarized
as follows. One strategy is to simplify the covariance structure of
GP, including low-rank GP \citep{cressie2008fixed,stein2008modeling},
inducing-point GP \citep{quinonero2005unifying,titsias2009variational},
local GP \citep{choudhury2002data,park2011domain}, covariance tapering
\citep{kaufman2008covariance,bevilacqua2016covariance}, and sparse
GP \citep{snelson2007local,sang2012full}. Another strategy is to
approximate the full likelihood of GP via the conditional composite
likelihood \citep{vecchia1988estimation,stein2004approximating,katzfuss2021general}
or block marginal composite likelihood \citep{caragea2007asymptotic,eidsvik2014estimation}.
All these approximation methods may be applied to PGP, but they will
inevitably suffer from a certain amount of information loss and thus
sacrifice some model accuracy for computational convenience. Moreover,
most of these methods are not scalable, and their computational complexity
still depends on the data size. Please refer to Section~\ref{sec:review}
for a thorough review.

In signal processing applications, data are often collected using
a given sampling frequency at grids. That is, the time series are
equally spaced. In dealing with such data, some algebraic methods
based on circulant and Toeplitz covariance structure may be used to
accelerate GP without any approximations. The embedding circulant
matrix \citep{wood1994simulation,coeurjolly2018fast} is a popular
method for fast and exact simulations from GPs. Yet, the covariance
matrix of PGP may not be circulant because the signal length $n$
may not always be multiples of the segment length $p$. So far, the
Toeplitz-based PGP (TPGP) is the most efficient method in the literature
that does not involve approximations in calculating likelihoods \citep{zhang2005time,chandola2011gaussian}.
It can accelerate the computational complexity of $\mathcal{O}\left(n^{3}\right)$
in PGP to $\mathcal{O}\left(n^{2}\right)$. Yet, this approach may
still be prohibitively slow when dealing with large-scale data.

To address the two challenges in classic PGP methods, we propose a
novel circulant PGP (CPGP) model for large-scale periodic data collected
at grids. The proposed CPGP divides the entire data into $k$ segments
and a remaining part if $n$ is not multiples of $k$. It decomposes
the log-likelihood of PGP into the sum of two computationally scalable
composite log-likelihoods, which accelerates the computational complexity
for calculating likelihoods from $\mathcal{O}\left(n^{3}\right)$
in PGP to $\mathcal{O}\left(p^{2}\right)$ in CPGP, where the segment
length $p$ is independent of and much smaller than the data size
$n$. When $p$ divides $n$, it can be further improved to $\mathcal{O}\left(p\log p\right)$.
Note that CPGP has exactly the same likelihoods as PGP given the period,
and thus it does not involve any approximations compared to PGP.

Different from classic PGP methods requiring known periods, the proposed
estimation of CPGP includes a tailored optimization algorithm that
can efficiently estimate periods if they are not known in advance.
Considering the cost of searching unknown periods in CPGP, the total
computational complexity for model fitting is $\mathcal{O}\left(d^{3}p_{max}^{3}\right)$,
where $p_{max}$ is the largest searching length for $p$ and $d$
is a tuning parameter controlling the decimal precision of period
estimations. Additionally, we also show that CPGP provides the exact
prediction as the best linear unbiased prediction (BLUP) in PGP, and
its computational complexity is $\mathcal{O}\left(p^{2}\right)$.
It clearly outperforms classic composite likelihood predictors that
may suffer from non-negligible information loss \citep{eidsvik2014estimation}.
Above all, the likelihood calculation, model fitting, and model prediction
in CPGP are all scalable, and no approximation is needed unlike those
for PGP.

When periodic data are collected at grids, the proposed CPGP should
be used wherever PGP models are applicable. It enables the use of
the accurate PGP framework to real applications that have large-scale
data or unknown periods. Notably, we do not advocate the use of CPGP
for data that are not collected at grids. Compared with some state-of-the-art
non-GP methods, including NLS \citep{nielsen2017fast}, NRC \citep{li2021extended},
and MLPE \citep{wise1976maximum}, CPGP performs better when dealing
with low SNR, such as vibration signals, and at the same time, is
computationally efficient. Besides the computational efficiency for
handling long signals, CPGP is particularly useful for fast and accurate
periodicity detection, i.e., accurately identifying the periods within
a very short time using a moderate number of data. Although CPGP is
designed for strictly periodic signals, it can also be used for analyzing
pseudo-periodic signals, which is illustrated in real case studies.
Additionally, we discuss an approximate version of CPGP in the simulation
study, which is faster but generally performs worse than CPGP.

The remainder of this article is organized as follows. In Section~\ref{sec:review},
we briefly review the current literature on GP, PGP, and some non-GP
methods for modeling periodic data. In Section~\ref{sec:SPGP}, we
detail the likelihood formulation, parameter estimation, and model
prediction of the scalable CPGP. In Section~\ref{sec:Simulation},
a simulation study on synthetic periodic signals is presented to evaluate
the performance of CPGP. In Section~\ref{sec:Case-Study}, two real
case studies are discussed to illustrate the superiority of CPGP.
In Section~\ref{sec:conc}, we conclude this study and discuss some
further work. All proofs and additional technical details are relegated
to Supplementary Materials.

\section{Brief Literature Review \label{sec:review} }

GP is a stochastic process where every finite collection of random
variables indexed by time or space follows a multivariate normal distribution.
The GP model, also known as Kriging, was first developed in the field
of geostatistics \citep{sacks1989designs} and then became widely
used in many areas of science and engineering \citep{kai2006design_book,gramacy2020surrogates}.
It can learn from temporal and/or spatial correlations and provide
desirable performance. Mathematically, the GP model (i.e. universal
Kriging) can be defined as 
\begin{equation}
y\left(t\right)=\boldsymbol{f}^{T}\left(t\right)\boldsymbol{\beta}+z\left(t\right)+\epsilon\left(t\right),\label{eq:GP0}
\end{equation}
where $\boldsymbol{f}\left(t\right)$ is a regression function, $\boldsymbol{\beta}$
is a vector of coefficients, $z\left(t\right)$ is a zero mean GP
with the variance $\sigma^{2}$ and stationary correlation function
$\varphi_{\boldsymbol{\phi}}(t,t^{'})$ ($\boldsymbol{\phi}$ is the
vector of parameters in the function), and $\epsilon\left(t\right)\sim\mathcal{N}\left(0,\sigma^{2}\delta^{2}\right)$
is an i.i.d. noise.

The parameters in the GP model ($\boldsymbol{\phi}$, $\boldsymbol{\beta}$,
$\sigma^{2}$, and $\delta^{2}$) are often obtained with maximum
likelihood estimation (MLE, \citealt{santner2013design_book}), i.e.,
by maximizing the following log-likelihood function (up to a constant)
\begin{equation}
\ell\left(\boldsymbol{\phi},\boldsymbol{\beta},\sigma^{2},\delta\right)=-\frac{\log\left|\sigma^{2}\boldsymbol{K}_{\delta}\right|+\frac{\left(\boldsymbol{y}-\boldsymbol{F}\boldsymbol{\beta}\right)^{T}\boldsymbol{K}_{\delta}^{-1}\left(\boldsymbol{y}-\boldsymbol{F}\boldsymbol{\beta}\right)}{\sigma^{2}}}{2},\label{eq:LIK0}
\end{equation}
where $\boldsymbol{F}=\left[\boldsymbol{f}\left(t_{1}\right),\cdots,\boldsymbol{f}\left(t_{n}\right)\right]^{T}$
is the regression matrix, and $\sigma^{2}\boldsymbol{K}_{\delta}$
is the covariance matrix of the GP. Here, the covariance $\boldsymbol{K}_{\delta}=\boldsymbol{K}+\delta^{2}\boldsymbol{I}_{n}$,
where the correlation matrix $\boldsymbol{K}$ is calculated from
$n$ sample points according to a chosen correlation function $\varphi_{\boldsymbol{\phi}}\left(t,t\right)$.

With the partial derivatives of the log-likelihood in Eq.~\eqref{eq:LIK0}
equated to zero, the parameter estimates of $\boldsymbol{\beta}$
and $\sigma^{2}$ are 
\begin{equation}
\begin{cases}
\hat{\boldsymbol{\beta}} & =\left(\boldsymbol{F}^{T}\boldsymbol{K}_{\delta}^{-1}\boldsymbol{F}\right)^{-1}\boldsymbol{F}^{T}\boldsymbol{K}_{\delta}^{-1}\boldsymbol{y}\\
\hat{\sigma}^{2} & =\frac{1}{n}\left(\boldsymbol{y}-\boldsymbol{F}\hat{\boldsymbol{\beta}}\right)^{T}\boldsymbol{K}_{\delta}^{-1}\left(\boldsymbol{y}-\boldsymbol{F}\hat{\boldsymbol{\beta}}\right)
\end{cases}.\label{eq:MLE0}
\end{equation}
Then, we can obtain the profile likelihood for the MLE of $\boldsymbol{\phi}$
and $\delta$, which is to solve the minimization problem $\hat{\boldsymbol{\phi}},\hat{\delta}=\mathrm{minimize}_{\boldsymbol{\phi},\delta}\left\{ n\log\hat{\sigma}^{2}+\log\left|\boldsymbol{K}_{\delta}\right|\right\} $
via some searching algorithms. Given parameter estimates, the model
prediction can be made through BLUP 
\begin{equation}
\hat{y}\left(t\right)=\boldsymbol{f}^{T}\left(t\right)\hat{\boldsymbol{\beta}}+\boldsymbol{r}^{T}\left(t\right)\boldsymbol{K}_{\hat{\delta}}^{-1}\left(\boldsymbol{y}-\boldsymbol{F}\hat{\boldsymbol{\beta}}\right),\label{eq:BLUP}
\end{equation}
where the correlation vector $\boldsymbol{r}\left(t\right)=\left[\varphi_{\boldsymbol{\hat{\phi}}}\left(t,t_{1}\right),\cdots,\varphi_{\boldsymbol{\hat{\phi}}}\left(t,t_{n}\right)\right]^{T}$.
Additionally, we have its variance as $\ensuremath{\mathrm{Var}\left(\hat{y}\left(t\right)\right)=\sigma^{2}(1-\boldsymbol{r}^{T}\left(t\right)\boldsymbol{K}_{\hat{\delta}}^{-1}\boldsymbol{r}\left(t\right)+\boldsymbol{\omega}^{T}\left(\boldsymbol{F}^{T}\boldsymbol{K}_{\delta}^{-1}\boldsymbol{F}\right)^{-1}\boldsymbol{\omega})}$,
where $\boldsymbol{\omega}=\boldsymbol{f}\left(t\right)-\boldsymbol{F}^{T}\boldsymbol{K}_{\hat{\delta}}^{-1}\boldsymbol{r}\left(t\right)$.
More details on GP derivations can be found in \citet{santner2013design_book}
and \citet{gramacy2020surrogates}.

Many usual correlation functions, e.g., Gaussian and Matern kernels,
may not be appropriate for modeling periodic data, because they do
not consider the periodic dependency. In the PGP literature, the periodic
correlation function proposed by \citet{mackay1998introduction} is
widely used, which is defined as 
\begin{equation}
\psi_{\boldsymbol{\phi}}\left(t,t'\right)=\exp\left(-\theta^{2}\sin^{2}\left(\frac{\pi\left(t-t'\right)}{T}\right)\right),\label{eq:cov_fun}
\end{equation}
where $\boldsymbol{\phi}=\left\{ \theta,T\right\} $, $\theta$ is
the roughness (length scale) parameter, and $T$ is the period. The
correlation function in Eq. \eqref{eq:cov_fun} can well model the
circulant correlation structure of periodic data, and the correlation
between the beginning and end points within any period converges to
one. More discussions and examples can be found in \citet{rasmussen2006gaussian}.

Model fitting and prediction for the GP model, including PGP, require
calculating the inverse and determinant of the covariance matrix,
both having a computational complexity of $\mathcal{O}\left(n^{3}\right)$.
For moderate or large datasets, it can be very time consuming. In
practice, periodic data, such as speech, ECG, and vibration signals,
often include more than 100,000 data points, and fitting a classic
GP or PGP is almost impossible for such data sizes.

To speed up GP methods, one common strategy is to simplify the covariance
structure via some approximations. For example, low-rank GP \citep{cressie2008fixed,stein2008modeling,finley2009hierarchical,xiong2021reconstruction}
was proposed to approximate the correlation matrix $\boldsymbol{K}$
by constructing the factorization $\boldsymbol{K}\approx\boldsymbol{Q}_{n\times p}\boldsymbol{K}_{p\times p}^{-1}\boldsymbol{Q}_{n\times p}^{T}$,
where $\boldsymbol{K}_{p\times p}$ is the correlation matrix of $p$
inducing points \citep{quinonero2005unifying,titsias2009variational}.
It can accelerate the computational complexity of GP to $\mathcal{O}\left(np^{2}\right)$.
Additionally, $\boldsymbol{K}_{p\times p}$ was rendered circulant
by \citet{tebbuttcircular} for pseudo-circular GP approximation,
which will further reduce the computational complexity to $\mathcal{O}\left(np\log p\right)$.
Moreover, local approximations \citep{choudhury2002data,park2011domain}
for large-scale GP were proposed by fitting local GP models on a subset
of data. Covariance tapering \citep{kaufman2008covariance,bevilacqua2016covariance}
was proposed by utilizing a sparse covariance structure for computational
convenience. Sparse GP \citep{snelson2007local,sang2012full} was
developed to combine the low-rank and local GP. Another strategy is
to apply divide-and-conquer with composite likelihoods. Specifically,
composite likelihoods were used to approximate the full likelihood
of GP for parameter estimation \citep{vecchia1988estimation,heagerty1998composite,stein2004approximating,caragea2007asymptotic}
and model prediction \citep{eidsvik2014estimation}. A sequential
pairwise modeling approach \citep{li2016pairwise} was developed to
achieve excellent scalability for multivariate GPs.

All these GP approximations can be applied to PGP, though most of
them have not been implemented in the current literature. One challenge
is that they may have non-negligible information loss due to approximations,
thereby affecting the accuracy of PGP. In signal processing applications,
periodic data are often collected at grids. For such data, the embedding
circulant matrix method \citep{wood1994simulation,davies2013circulant}
can be used to scale GP without any approximations, where the Toeplitz
covariance matrix of GP can be embedded into a circulant covariance
matrix. Calculating the inverse and determinant of the covariance
matrix requires $\mathcal{O}\left(n\log n\right)$ and $\mathcal{O}\left(n^{2}\right)$
time, respectively, and thus the complexity of computing likelihoods
is $\mathcal{O}\left(n^{2}\right)$ \citep{zhang2005time,chandola2011gaussian}.
This method has been successfully applied to GP simulations \citep{dietrich1997fast,coeurjolly2018fast}
and log-Gaussian Cox processes \citep{diggle2013spatial,taylor2014inla}.

Specifically, TPGP proposed by \citet{chandola2011gaussian} is currently
the most efficient PGP dealing with periodic data collected at grids,
i.e., $t_{i}=i/f_{s}$, where $f_{s}$ is the sampling frequency.
Its covariance matrix $\sigma^{2}\boldsymbol{K}_{\delta}$ is proved
to be a Toeplitz matrix (diagonal-constant matrix) for any period
$T$, because any $i$th and $j$th element of the correlation matrix
$\boldsymbol{K}$ is 
\begin{equation}
\psi_{\boldsymbol{\phi}}\left(t_{i},t_{j}\right)=\exp\left(-\theta^{2}\sin^{2}\left(\frac{\pi\left(i/f_{s}-j/f_{s}\right)}{T}\right)\right).\label{eq:cov_fun1}
\end{equation}
Clearly, all descending diagonals of $\boldsymbol{K}$ are constant,
and so is $\boldsymbol{K}_{\delta}=\boldsymbol{K}+\delta^{2}\boldsymbol{I}_{n}$.
Then, a Toeplitz matrix inversion algorithm \citep{trench1964algorithm}
can be used to accelerate the computation of likelihoods in $\mathcal{O}\left(n^{2}\right)$
time.

The proposed CPGP differs from current GP (or PGP) methods in the
following four aspects. First, different from all GP (or PGP) approximation
methods, CPGP has exactly the same likelihood as PGP for periodic
data collected at grids, thus maintaining the same accuracy. Second,
CPGP is a scalable approach that requires only $\mathcal{O}\left(p^{2}\right)$
time (or $\mathcal{O}\left(p\log p\right)$ in a special case) for
calculating likelihoods, where $p$ is independent of and much smaller
than $n$. In contrast, most current GP methods are still not scalable
even after acceleration, e.g., low rank GP and TPGP have a computational
complexity of $\mathcal{O}\left(np\log p\right)$ and of $\mathcal{O}\left(n^{2}\right)$,
respectively. Third, CPGP can handle the case where $n$ is not multiples
of $p$, while current methods based on circulant matrices cannot,
though they all leverage fast Fourier transform (FFT) to accelerate
the computation. Fourth, unlike current PGP methods that commonly
assume known periods, the estimation of CPGP includes an efficient
optimization for identifying unknown periods. Note that besides GP-based
methods, we will also compare the proposed CPGP with some state-of-the-art
non-GP methods, including fast NLS (FNLS, \citealt{nielsen2017fast}),
NRC \citep{li2021extended}, and MLPE \citep{wise1976maximum}, via
numerical studies.

\section{Circulant Periodic Gaussian Process \label{sec:SPGP}}

Large-scale periodic data are often collected at grids by using a
given sampling frequency in signal processing applications. To enable
an accurate PGP framework for modeling such data, we propose a scalable
modeling approach, called CPGP, which is fast in both parameter estimation
and model prediction. Following the settings in \citet{zhang2005time}
and \citet{chandola2011gaussian}, we assume the periodic data are
collected at grids with the sampling frequency $f_{s}$ and the period
can be written as $T=p/\left(df_{s}\right)$, where $p$ and $d$
are co-prime integers. The tuning parameter $d$ is used to allow
a decimal number of points (i.e., $p/d$ points) in one period and
control the estimation accuracy of $T$. In practice, the sampling
frequency $f_{s}$ for collecting data is often given, while the true
period $T$ of the signal is often unknown. By plugging $T=p/\left(df_{s}\right)$
into Eq. \eqref{eq:cov_fun1}, the correlation function $\psi_{\boldsymbol{\phi}}\left(t_{i},t_{j}\right)$
can be written as 
\begin{equation}
\psi_{\boldsymbol{\phi}}\left(t_{i},t_{j}\right)=\exp\left(-\theta^{2}\sin^{2}\left(\frac{\pi d\left(i-j\right)}{p}\right)\right),\label{eq:corr_psi}
\end{equation}
where $i,j=1,\ldots,n$ and $\boldsymbol{\phi}=\left\{ \theta,p,d\right\} $.

\subsection{Exact Likelihood Decomposition}

\label{sec:ld} In CPGP, we divide the periodic signal collected at
grids into $k=\left\lfloor n/p\right\rfloor $ segments, each having
$p$ data points, where $\left\lfloor n/p\right\rfloor $ denotes
the maximum integer not greater than $n/p$. The $r$th segment is
denoted as $\boldsymbol{y}_{r}=\left[y\left(t_{\left(r-1\right)p+1}\right),...,y\left(t_{rp}\right)\right]^{T}$
for $r=1,2,\cdots,k$, and the remaining data points that cannot form
a segment are denoted as $\boldsymbol{y}_{*}=\left[y\left(t_{kp+1}\right),...,y\left(t_{n}\right)\right]^{T}$.

A key novelty of CPGP is that it decomposes the log-likelihood $\ell$
in PGP into the sum of two computationally scalable composite log-likelihoods
without any approximations. Specifically, by applying the conditional
probability, we have 
\begin{equation}
\ell\left(\theta,\delta,p,d\right)=\ell_{1}+\eta\ell_{2},\label{eq:log-lik}
\end{equation}
where $\ell_{1}=\log\mathsf{Pr}\left(\boldsymbol{y}_{1},\cdots,\boldsymbol{y}_{k}\right)$
is the log-likelihood of $\boldsymbol{y}_{1},\cdots,\boldsymbol{y}_{k}$,
$\ell_{2}=\log\mathsf{Pr}\left(\boldsymbol{y}_{*}\mid\boldsymbol{y}_{1},\cdots,\boldsymbol{y}_{k}\right)$
is the log-likelihood of $\boldsymbol{y}_{*}$ conditional on $\boldsymbol{y}_{1},\cdots,\boldsymbol{y}_{k}$,
and $\eta=0$ if $n\mod p=0$, otherwise $\eta=1$. Notably, $\ell_{2}$
should not be ignored here, as CPGP is an exact version (rather than
an approximation) of PGP. On the one hand, $\ell_{2}$ is needed to
make $\ell_{1}$ comparable among different sizes of $p$, as the
sample size in $\ell_{1}$ changes with $p$. On the other hand, dropping
$\ell_{2}$ may lead to worse performance, as illustrated in the following
example, as well as the simulation study in Section~\ref{sec:Simulation}
where an approximate CPGP is discussed.
\begin{example}
\label{eg:l2} Consider the synthetic periodic signal in \citet{fan2018noise}
with data size $n=4,000$, a sampling frequency $f_{s}=1$ Hz, and
an SNR of $-18$ dB; refer to Section~\ref{sec:Simulation} for details.
We fit a classic PGP in Eq. \eqref{eq:GP0} using the correlation
function in Eq. \eqref{eq:corr_psi} with parameters $\theta=15$,
$\delta=3$ and $d=1$. We show the waveforms of its full log-likelihood
$\ell$ and its first composite log-likelihood $\ell_{1}$ changing
with $p$ in Fig. \ref{fig:likelihood}. It is seen that $\ell_{1}$
fluctuates dramatically as $p$ changes, because a changing signal
length (i.e., $kp$) is used in $\ell_{1}$. In the bottom panel of
Fig. \ref{fig:likelihood}, we further show the normalized $\ell_{1}$
(i.e., $\ell_{1}/\left(kp\right)$), which still fluctuates considerably.
The highest value of the full likelihood $\ell$ locates at $p=200$
(i.e., the true period). Yet, the highest value of $\ell_{1}$ or
$\ell_{1}/\left(kp\right)$ locates at a much larger $p$. Clearly,
the second composite log-likelihood $\ell_{2}$ should not be ignored
here. 
\begin{figure}[t]
\begin{centering}
\includegraphics[width=9.5cm]{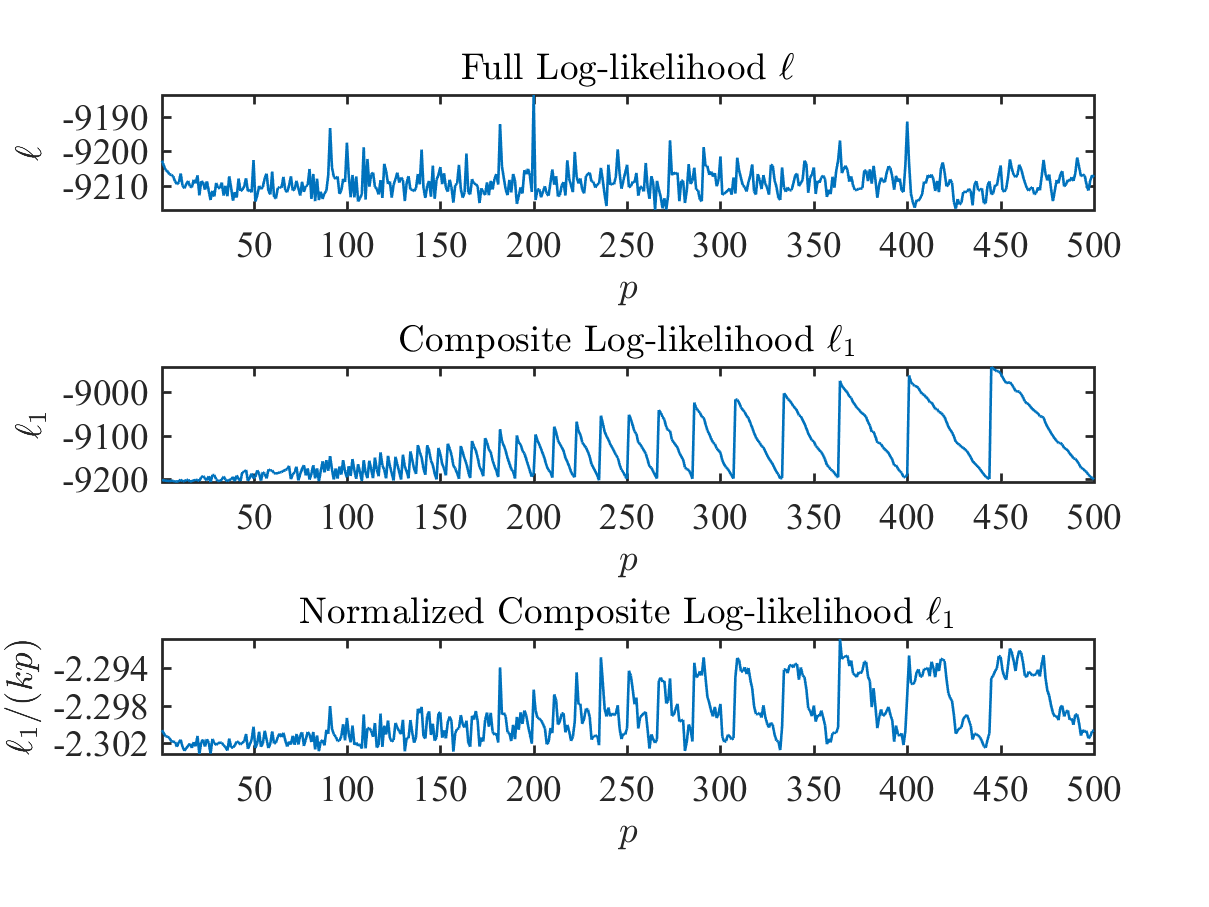} 
\par\end{centering}
\caption{\label{fig:likelihood}Values of $\ell$ and $\ell_{1}$ changing
with $p$.}
\end{figure}
\end{example}
After the signal is divided into $k$ segments $\boldsymbol{y}_{r}$
($r=1,2,\cdots,k$) and the remaining $\boldsymbol{y}_{*}$, the covariance
matrix between $\boldsymbol{y}_{r}$ and $\boldsymbol{y}_{s}$ is
$\sigma^{2}\left(\boldsymbol{R}+\delta^{2}\boldsymbol{I}_{p}\right)$
if $r=s$, and $\sigma^{2}\boldsymbol{R}$ if $r\neq s$, where 
\[
\boldsymbol{R}=\left[\begin{array}{ccc}
\psi_{\boldsymbol{\phi}}\left(t_{1},t_{1}\right) & \cdots & \psi_{\boldsymbol{\phi}}\left(t_{1},t_{p}\right)\\
\vdots & \ddots & \vdots\\
\psi_{\boldsymbol{\phi}}\left(t_{p},t_{1}\right) & \cdots & \psi_{\boldsymbol{\phi}}\left(t_{p},t_{p}\right)
\end{array}\right];
\]
the covariance matrix between $\boldsymbol{y}_{r}$ and $\boldsymbol{y}_{*}$
is $\sigma^{2}\boldsymbol{R}_{\bullet}$, where 
\[
\boldsymbol{R}_{\bullet}=\left[\begin{array}{ccc}
\psi_{\boldsymbol{\phi}}\left(t_{1},t_{1}\right) & \cdots & \psi_{\boldsymbol{\phi}}\left(t_{1},t_{n-kp}\right)\\
\vdots & \ddots & \vdots\\
\psi_{\boldsymbol{\phi}}\left(t_{p},t_{1}\right) & \cdots & \psi_{\boldsymbol{\phi}}\left(t_{p},t_{n-kp}\right)
\end{array}\right];
\]
the covariance matrix for $\boldsymbol{y}_{*}$ is $\sigma^{2}\left(\boldsymbol{R}_{*}+\delta^{2}\boldsymbol{I}_{n-kp}\right)$,
where 
\[
\boldsymbol{R}_{*}=\left[\begin{array}{ccc}
\psi_{\boldsymbol{\phi}}\left(t_{1},t_{1}\right) & \cdots & \psi_{\boldsymbol{\phi}}\left(t_{1},t_{n-kp}\right)\\
\vdots & \ddots & \vdots\\
\psi_{\boldsymbol{\phi}}\left(t_{n-kp},t_{1}\right) & \cdots & \psi_{\boldsymbol{\phi}}\left(t_{n-kp},t_{n-kp}\right)
\end{array}\right].
\]

Next, we obtain a key result for the decomposition of correlation
matrix $\boldsymbol{K}$ and the properties of matrices $\boldsymbol{R}$
and $\boldsymbol{R}_{*}$. 
\begin{prop}
\label{prop1} The correlation matrix $\boldsymbol{K}$ by Eq. \eqref{eq:corr_psi}
can be decomposed as 
\begin{equation}
\boldsymbol{K}=\left[\begin{array}{cccc}
\boldsymbol{R} & \cdots & \boldsymbol{R} & \boldsymbol{R}_{\bullet}\\
\vdots & \ddots & \vdots & \vdots\\
\boldsymbol{R} & \cdots & \boldsymbol{R} & \boldsymbol{R}_{\bullet}\\
\boldsymbol{R}_{\bullet}^{T} & \cdots & \boldsymbol{R}_{\bullet}^{T} & \boldsymbol{R}_{*}
\end{array}\right],\label{eq:decomposition}
\end{equation}
where $\boldsymbol{R}$ is a symmetric circulant matrix and $\boldsymbol{R}_{*}$
is a symmetric Toeplitz matrix. 
\end{prop}
The joint distribution of segments $\boldsymbol{\varUpsilon}=\left[\boldsymbol{y}_{1}^{T},\cdots,\boldsymbol{y}_{k}^{T}\right]^{T}$
is 
\begin{align*}
\boldsymbol{\varUpsilon} & \sim\mathcal{N}\left(\boldsymbol{\varGamma}\boldsymbol{\beta},\sigma^{2}\boldsymbol{\Sigma}\right)\\
 & \sim\mathcal{N}\left(\left[\begin{array}{c}
\boldsymbol{\varGamma}_{1}\boldsymbol{\beta}\\
\vdots\\
\boldsymbol{\varGamma}_{k}\boldsymbol{\beta}
\end{array}\right],\sigma^{2}\left[\begin{array}{ccc}
\boldsymbol{R}+\delta^{2}\boldsymbol{I}_{p} & \cdots & \boldsymbol{R}\\
\vdots & \ddots & \vdots\\
\boldsymbol{R} & \cdots & \boldsymbol{R}+\delta^{2}\boldsymbol{I}_{p}
\end{array}\right]\right),
\end{align*}
where $\boldsymbol{\varGamma}_{i}=\left[\boldsymbol{f}\left(t_{1+\left(i-1\right)p}\right),\cdots,\boldsymbol{f}\left(t_{ip}\right)\right]^{T}$
for $i=1,\cdots,k$. Then, the first composite log-likelihood $\ell_{1}$,
up to a constant, can be written as 
\begin{equation}
\ell_{1}=-\frac{1}{2}\left(\log\left|\sigma^{2}\boldsymbol{\Sigma}\right|+\frac{\left(\boldsymbol{\varUpsilon}-\boldsymbol{\varGamma}\boldsymbol{\beta}\right)^{T}\boldsymbol{\Sigma}^{-1}\left(\boldsymbol{\varUpsilon}-\boldsymbol{\varGamma}\boldsymbol{\beta}\right)}{\sigma^{2}}\right).\label{eq:log-lik1}
\end{equation}

The distribution of $\boldsymbol{y}_{*}$, conditional on $\boldsymbol{y}_{1},\cdots,\boldsymbol{y}_{k}$,
is $\boldsymbol{y}_{*}\mid\boldsymbol{\varUpsilon}\sim\mathcal{N}\left(\boldsymbol{\mu},\sigma^{2}\boldsymbol{\varPi}\right)$
with the mean $\boldsymbol{\mu}=\boldsymbol{\varGamma}_{*}\boldsymbol{\beta}+\boldsymbol{\Xi}^{T}\boldsymbol{\Sigma}^{-1}\left(\boldsymbol{\varUpsilon}-\boldsymbol{\varGamma}\boldsymbol{\beta}\right)$
and the covariance matrix $\boldsymbol{\varPi}=\boldsymbol{R}_{*}+\delta^{2}\boldsymbol{I}_{n-kp}-\boldsymbol{\Xi}^{T}\boldsymbol{\Sigma}^{-1}\boldsymbol{\Xi}$,
where $\boldsymbol{\Xi}=\left[\boldsymbol{R}_{\bullet}^{T}\cdots\boldsymbol{R}_{\bullet}^{T}\right]^{T}$
and $\boldsymbol{\varGamma}_{*}=\left[\boldsymbol{f}\left(t_{kp+1}\right),\cdots,\boldsymbol{f}\left(t_{n}\right)\right]^{T}$.
Then, the second composite log-likelihood $\ell_{2}$, up to some
constants, can be written as 
\begin{equation}
\ell_{2}=-\frac{1}{2}\left(\log\left|\sigma^{2}\boldsymbol{\varPi}\right|+\frac{\left(\boldsymbol{y}_{\bullet}-\boldsymbol{\varGamma}_{\bullet}\boldsymbol{\beta}\right)^{T}\boldsymbol{\varPi}^{-1}\left(\boldsymbol{y}_{\bullet}-\boldsymbol{\varGamma}_{\bullet}\boldsymbol{\beta}\right)}{\sigma^{2}}\right),\label{eq:log-lik2}
\end{equation}
where $\boldsymbol{y}_{\bullet}=\boldsymbol{y}_{*}-\boldsymbol{\Xi}^{T}\boldsymbol{\Sigma}^{-1}\boldsymbol{\varUpsilon}$
and $\boldsymbol{\varGamma}_{\bullet}=\boldsymbol{\varGamma}_{*}-\boldsymbol{\Xi}^{T}\boldsymbol{\Sigma}^{-1}\boldsymbol{\varGamma}$.
Note that if $p>n$, then the composite log-likelihood $\ell_{1}$
reduces to zero, and the composite log-likelihood $\ell_{2}$ reduces
to $\ell_{2}=-1/2(\log\left|\sigma^{2}\boldsymbol{R}_{*}\right|+\left(\boldsymbol{y}_{*}-\boldsymbol{\varGamma}_{*}\boldsymbol{\beta}\right)^{T}\boldsymbol{R}_{*}^{-1}\left(\boldsymbol{y}_{*}-\boldsymbol{\varGamma}_{*}\boldsymbol{\beta}\right)/\sigma^{2})$.

\subsection{Scalable Parameter Estimation}

In accordance with Eqs. \eqref{eq:log-lik1} and \eqref{eq:log-lik2},
the assessments of $\ell_{1}$ and $\ell_{2}$ require the determinant
and inverse of $\boldsymbol{\boldsymbol{\Sigma}}$ and $\boldsymbol{\varPi}$,
respectively. Since $\boldsymbol{\varPi}=\boldsymbol{R}_{*}+\delta^{2}\boldsymbol{I}_{n-kp}-\boldsymbol{\Xi}^{T}\boldsymbol{\Sigma}^{-1}\boldsymbol{\Xi}$,
the key challenge lies in the calculation of $\left|\boldsymbol{\boldsymbol{\Sigma}}\right|$
and $\boldsymbol{\boldsymbol{\Sigma}}^{-1}$. In Proposition~\ref{prop2},
we show that they can be substantially simplified by utilizing the
matrix inversion lemma \citep{sherman1950adjustment} and Sylvester
determinant theorem \citep{akritas1996various}. The matrix inversion
lemma is also known as the Woodbury, Sherman, and Morrison formula. 
\begin{prop}
\label{prop2} The inverse and determinant of $\boldsymbol{\Sigma}$
can be simplified as 
\begin{equation}
\boldsymbol{\Sigma}^{-1}=\frac{k\boldsymbol{I}_{kp}-\boldsymbol{V}\left(\boldsymbol{I}_{p}-\boldsymbol{R}_{\delta}^{-1}\right)\boldsymbol{V}^{T}}{\delta^{2}k},\label{eq:Inv}
\end{equation}
and 
\begin{equation}
\left|\boldsymbol{\Sigma}\right|=\delta^{2kp}\left|\boldsymbol{R}_{\delta}\right|,\label{eq:Det}
\end{equation}
where $\boldsymbol{R}_{\delta}=\boldsymbol{I}_{p}+k\boldsymbol{R}/\delta^{2}$
and $\boldsymbol{V}=\left[\begin{array}{ccc}
\boldsymbol{I}_{p} & \cdots & \boldsymbol{I}_{p}\end{array}\right]^{T}$. 
\end{prop}
According to Proposition~\ref{prop1}, the matrix $\boldsymbol{R}$
is a symmetric circulant matrix. Thus, the matrices $\boldsymbol{R}_{\delta}$
and $\boldsymbol{R}_{\delta}^{-1}$ in Proposition~\ref{prop2} are
symmetric circulant matrices. Then, FFT method \citep{brigham1988fast}
can be applied to compute $\boldsymbol{R}_{\delta}^{-1}$ and $\left|\boldsymbol{R}_{\delta}\right|$
with a computational complexity of $\mathcal{O}\left(p\log p\right)$.
The MATLAB toolbox ``smt\textquotedbl{} \citep{redivo2012smt} can
be used for implementation.

By plugging Eq. \eqref{eq:Inv} into the formulas of $\boldsymbol{y}_{\bullet}$,
$\boldsymbol{\varGamma}_{\bullet}$, and $\boldsymbol{\varPi}$ defined
above, we have 
\[
\begin{cases}
\boldsymbol{y}_{\bullet} & =\boldsymbol{y}_{*}-\frac{k}{\delta^{2}}\boldsymbol{R}_{\bullet}^{T}\boldsymbol{R}_{\delta}^{-1}\bar{\boldsymbol{y}}\\
\boldsymbol{\varGamma}_{\bullet} & =\boldsymbol{\varGamma}_{*}-\frac{k}{\delta^{2}}\boldsymbol{R}_{\bullet}^{T}\boldsymbol{R}_{\delta}^{-1}\bar{\boldsymbol{\varGamma}}\\
\boldsymbol{\varPi} & =\boldsymbol{R}_{*}+\delta^{2}\boldsymbol{I}_{n-kp}-\frac{k}{\delta^{2}}\boldsymbol{R}_{\bullet}^{T}\boldsymbol{R}_{\delta}^{-1}\boldsymbol{R}_{\bullet}
\end{cases},
\]
where $\bar{\boldsymbol{y}}=\left(\boldsymbol{y}_{1}+\cdots+\boldsymbol{y}_{k}\right)/k$
and $\bar{\boldsymbol{\varGamma}}=\left(\boldsymbol{\varGamma}_{1}+\cdots+\boldsymbol{\varGamma}_{k}\right)/k$.
Here, $\boldsymbol{R}^{T}\boldsymbol{R}_{\delta}^{-1}\boldsymbol{R}$
is a symmetric circulant matrix because the product of circulant matrices
is still circulant. The matrix $\boldsymbol{R}_{\bullet}^{T}\boldsymbol{R}_{\delta}^{-1}\boldsymbol{R}_{\bullet}$
(a sub-matrix of $\boldsymbol{R}^{T}\boldsymbol{R}_{\delta}^{-1}\boldsymbol{R}$)
is a symmetric Toeplitz matrix, and so is $\boldsymbol{\varPi}$.
Then, $\boldsymbol{\varPi}$ can be calculated in $\mathcal{O}\left(p\log p\right)$
time because we only need to compute the first row/column of $\boldsymbol{\varPi}$
via FFT. We can use the Cholesky factorization algorithm for semidefinite
Toeplitz matrices \citep{stewart1997cholesky} to calculate $\boldsymbol{\varPi}^{-1}$
and $\left|\boldsymbol{\varPi}\right|$, which requires $\mathcal{O}\left(\left(n-kp\right)^{2}\right)$
time. It is straightforward to show that this complexity is always
smaller than $\mathcal{O}\left(p^{2}\right)$. Note that we may further
accelerate the calculation of $\boldsymbol{\varPi}^{-1}$ in $\mathcal{O}\left(p\log p\right)$
time by embedding $\boldsymbol{\varPi}$ into some circulant matrix,
but the calculation of $\left|\boldsymbol{\varPi}\right|$ cannot
be further improved. Thus, the embedding circulant matrix method cannot
further improve the complexity for calculating $\ell_{2}$ (involving
both $\boldsymbol{\varPi}^{-1}$ and $\left|\boldsymbol{\varPi}\right|$).
Here, we consider only the Cholesky factorization algorithm.

By solving $\partial\ell/\partial\boldsymbol{\beta}=0$ and $\partial\ell/\partial\sigma^{2}=0$,
we obtain the estimations of $\boldsymbol{\beta}$ and $\sigma^{2}$
\begin{equation}
\hat{\boldsymbol{\beta}}=\left(\boldsymbol{S}_{\boldsymbol{\varGamma}\boldsymbol{\varGamma}}+\eta\boldsymbol{\varGamma}_{\bullet}^{T}\boldsymbol{\varPi}^{-1}\boldsymbol{\varGamma}_{\bullet}\right)^{-1}\left(\boldsymbol{S}_{\boldsymbol{\varGamma}\boldsymbol{Y}}+\eta\boldsymbol{\varGamma}_{\bullet}^{T}\boldsymbol{\varPi}^{-1}\boldsymbol{y}_{\bullet}\right),\label{eq:beta_hat}
\end{equation}
and 
\begin{equation}
\hat{\sigma}^{2}=\frac{\left(\boldsymbol{S}_{\boldsymbol{Y}\boldsymbol{Y}}+\eta\boldsymbol{y}_{\bullet}^{T}\boldsymbol{\varPi}^{-1}\boldsymbol{y}_{\bullet}\right)-\hat{\boldsymbol{\beta}}^{T}\left(\boldsymbol{S}_{\boldsymbol{\varGamma}\boldsymbol{\varGamma}}+\eta\boldsymbol{\varGamma}_{\bullet}^{T}\boldsymbol{\varPi}^{-1}\boldsymbol{\varGamma}_{\bullet}\right)\hat{\boldsymbol{\beta}}}{n},\label{eq:sigma_hat}
\end{equation}
where 
\[
\begin{cases}
\boldsymbol{S}_{\boldsymbol{\varGamma}\boldsymbol{\varGamma}} & =\boldsymbol{\varGamma}^{T}\boldsymbol{\boldsymbol{\Sigma}}^{-1}\boldsymbol{\varGamma}=\frac{\boldsymbol{\varGamma}^{T}\boldsymbol{\varGamma}-k\bar{\boldsymbol{\varGamma}}^{T}\bar{\boldsymbol{\varGamma}}+k\bar{\boldsymbol{\varGamma}}^{T}\boldsymbol{R}_{\delta}^{-1}\bar{\boldsymbol{\varGamma}}}{\delta^{2}}\\
\boldsymbol{S}_{\boldsymbol{\varGamma}\boldsymbol{Y}} & =\boldsymbol{\varGamma}^{T}\boldsymbol{\boldsymbol{\Sigma}}^{-1}\boldsymbol{Y}=\frac{\boldsymbol{\varGamma}^{T}\boldsymbol{Y}-k\bar{\boldsymbol{\varGamma}}^{T}\bar{\boldsymbol{y}}+k\bar{\boldsymbol{\varGamma}}^{T}\boldsymbol{R}_{\delta}^{-1}\bar{\boldsymbol{y}}}{\delta^{2}}\\
\boldsymbol{S}_{\boldsymbol{Y}\boldsymbol{Y}} & =\boldsymbol{Y}^{T}\boldsymbol{\boldsymbol{\Sigma}}^{-1}\boldsymbol{Y}=\frac{\boldsymbol{Y}^{T}\boldsymbol{Y}-k\bar{\boldsymbol{y}}^{T}\bar{\boldsymbol{y}}+k\bar{\boldsymbol{y}}^{T}\boldsymbol{R}_{\delta}^{-1}\bar{\boldsymbol{y}}}{\delta^{2}}
\end{cases}.
\]
Specifically, if $\boldsymbol{\varGamma}_{1}=\cdots=\boldsymbol{\varGamma}_{k}$,
$\boldsymbol{S}_{\boldsymbol{\varGamma}\boldsymbol{\varGamma}}=k\bar{\boldsymbol{\varGamma}}^{T}\boldsymbol{R}_{\delta}^{-1}\bar{\boldsymbol{\varGamma}}/\delta^{2}$
and $\boldsymbol{S}_{\boldsymbol{\varGamma}\boldsymbol{Y}}=k\bar{\boldsymbol{\varGamma}}^{T}\boldsymbol{R}_{\delta}^{-1}\bar{\boldsymbol{y}}/\delta^{2}$.
Here, calculating $\hat{\boldsymbol{\beta}}$ and $\hat{\sigma}^{2}$
involves the calculation of $\boldsymbol{R}_{\delta}^{-1}$ and $\boldsymbol{\varPi}^{-1}$
whose dimensionalities are no more than $p$. This shows that the
estimations of $\hat{\boldsymbol{\beta}}$ and $\hat{\sigma}^{2}$
are scalable.

By plugging Eqs. \eqref{eq:Det}, \eqref{eq:beta_hat}, and \eqref{eq:sigma_hat}
into Eqs. \eqref{eq:log-lik1} and \eqref{eq:log-lik2}, the log-likelihood
$\ell\left(\theta,\delta,p,d\right)$ in Eq. \eqref{eq:log-lik} can
be written as 
\begin{equation}
\hat{\ell}\left(\theta,\delta,p,d\right)=-\frac{\log\hat{\sigma}^{2n}\delta^{2kp}+\log\left|\boldsymbol{R}_{\delta}\right|\left|\boldsymbol{\varPi}\right|^{\eta}+n}{2}.\label{loglik}
\end{equation}
Here, the evaluation of $\hat{\ell}\left(\theta,\delta,p,d\right)$
involves calculating $\boldsymbol{R}_{\delta}$ and $\boldsymbol{\varPi}$,
which requires $\mathcal{O}\left(p\log p\right)$ and $\mathcal{O}\left(p^{2}\right)$
time, respectively. Thus, calculating the likelihoods of CPGP requires
a computational complexity of $\mathcal{O}\left(p^{2}\right)$ in
general. When $p$ divides $n$ ($\ell=\ell_{1}$ and $\ell_{2}=0$),
the evaluation of $\hat{\ell}\left(\theta,\delta,p,d\right)$ involves
calculating only $\boldsymbol{\Sigma}^{-1}$ and $\vert\boldsymbol{\Sigma}\vert$.
According to Proposition~\ref{prop2}, such a special case requires
only $\mathcal{O}\left(p\log p\right)$ time. As a comparison, the
PGP and TPGP have a complexity of $\mathcal{O}\left(n^{3}\right)$
and $\mathcal{O}\left(n^{2}\right)$, respectively. Next, we give
an example to illustrate the speeds of PGP, TPGP, and CPGP for calculating
likelihoods. 
\begin{example}
Consider the synthetic periodic signals in Example~\ref{eg:l2} with
its length $n$ ranging from 1,000 to 500,000. The PGP, TPGP, and
CPGP models are implemented to evaluate the log-likelihood $\hat{\ell}$
with parameters $\theta=15$ and $\delta=10$. The average computing
time of 100 trials is reported in Fig. \ref{fig:compTime}. Unlike
PGP and TPGP, the proposed CPGP has a computing time independent of
the signal length. Moreover, CPGP is substantially faster than PGP
and TPGP for long signals. In addition, we can see that CPGP with
a predetermined period $T=100$ (i.e., $p=100$ and $d=1$) is faster
than that with $T=11$ (i.e., $p=11$ and $d=1$), because its complexity
improves from $\mathcal{O}\left(p^{2}\right)$ in general to $\mathcal{O}\left(p\log p\right)$
when $p$ divides $n$. CPGP can also handle decimal periods, as seen
in the cases of $T=30.1$ ($p=301$ and $d=10$) and $T=80.1$ ($p=801$
and $d=10$). 
\begin{figure}[t]
\begin{centering}
\includegraphics[width=9.5cm]{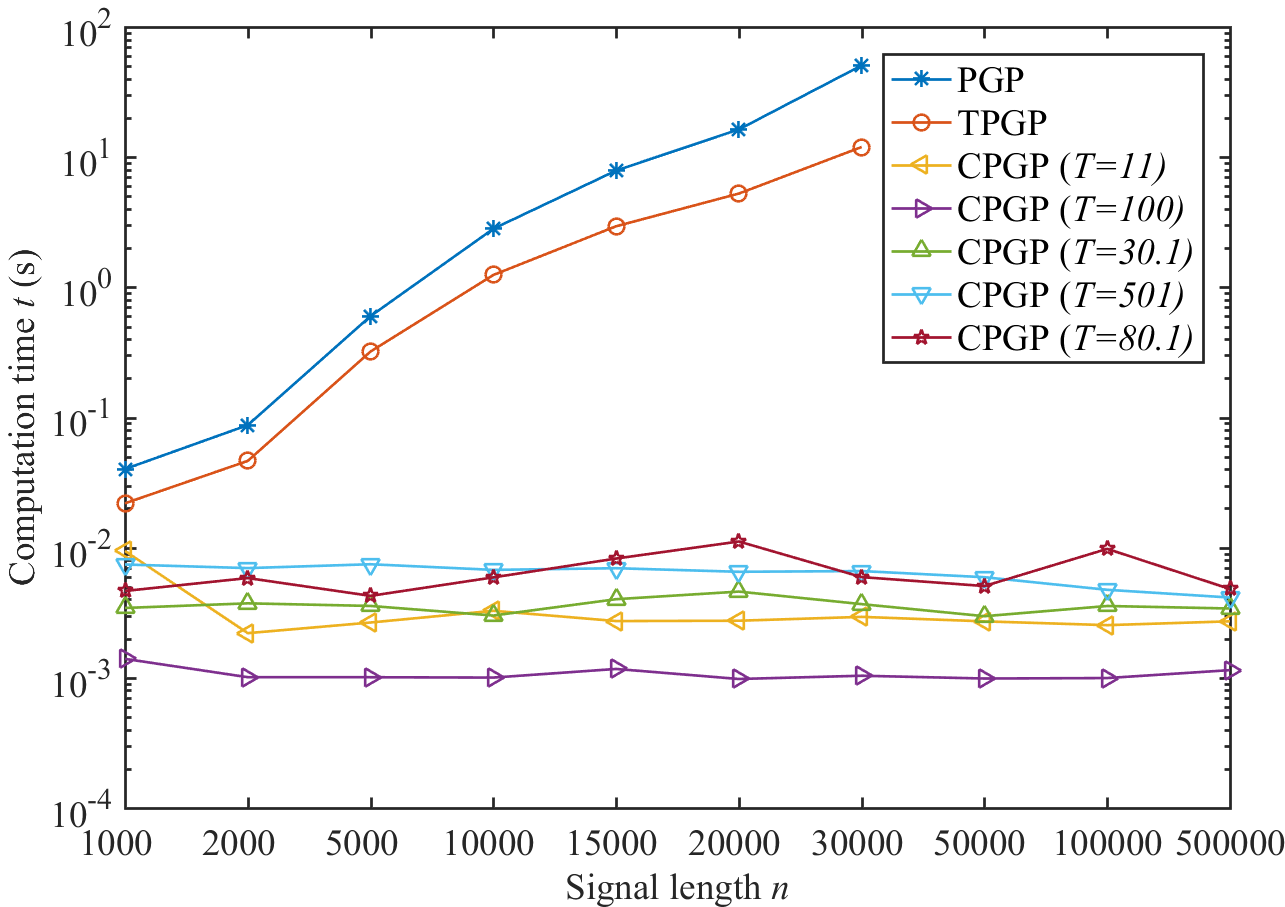} 
\par\end{centering}
\caption{\label{fig:compTime}Average computational time of PGP, TPGP, and
CPGP.}
\end{figure}
\end{example}
In practice, the sampling frequency $f_{s}$ for collecting data is
usually specified in advance, while the period $T=p/\left(df_{s}\right)$
is often unknown (and so is the segment length $p$). There are often
numerous local optimums in likelihoods with regard to $p$, as shown
in Fig.~\ref{fig:likelihood}, and thus estimating $p$ is non-trivial.
In this work, we propose to locate the maximum value of $\hat{\ell}\left(\theta,\delta,p,d_{*}\right)$
in Eq.~\eqref{loglik} by scanning $p$ in $\boldsymbol{\mathcal{I}}=\left\{ 1,2,\cdots,d_{*}p_{max}\right\} $
before a search algorithm is applied to estimate $\theta$ and $\delta^{2}$,
where $d_{*}$ is a tuning parameter such that $1\leq d_{*}\leq d$
for computational saving. That is, we will first locate the maximum
log-likelihood $\hat{\ell}_{\boldsymbol{\mathcal{I}}}\left(\theta,\delta,d_{*}\right)=\max_{p\in\boldsymbol{\mathcal{I}}}\hat{\ell}\left(\theta,\delta,p,d_{*}\right)$
by scanning $p$ in $\boldsymbol{\mathcal{I}}$ and then optimize
$\hat{\ell}_{\boldsymbol{\mathcal{I}}}\left(\theta,\delta,d_{*}\right)$
to estimate $\theta$ and $\delta^{2}$ via some searching algorithms,
i.e., 
\begin{equation}
\hat{\theta},\hat{\delta}=\mathrm{maximize}_{\theta,\delta}\left\{ \hat{\ell}_{\boldsymbol{\mathcal{I}}}\left(\theta,\delta,d_{*}\right)\right\} .\label{eq:OPT0}
\end{equation}
Many heuristic optimization algorithms, such as gradient descent algorithms,
can be used here, and we adopt the Hooke-Jeeves searching algorithm
implemented by the MATLAB toolbox ``dace'' \citep{lophaven2002aspects}.

By plugging the optimized $\hat{\theta}$ and $\hat{\delta}$ from
Eq.~\eqref{eq:OPT0} into $\hat{\ell}\left(\theta,\delta,p,d\right)$
in Eq. \eqref{loglik}, we have (given $p$ and $d$) 
\begin{equation}
\hat{\ell}\left(\hat{\theta},\hat{\delta},p,d\right)=-\frac{\log\hat{\sigma}^{2n}\hat{\delta}^{2kp}+\log\left|\hat{\boldsymbol{R}}_{\hat{\delta}}\right|\left|\hat{\boldsymbol{\varPi}}\right|^{\eta}+n}{2},\label{loglik1}
\end{equation}
where $\hat{\boldsymbol{R}}_{\hat{\delta}}$ and $\hat{\boldsymbol{\varPi}}$
can be obtained by plugging $\hat{\theta}$ and $\hat{\delta}$ into
$\boldsymbol{R}_{\delta}$ and $\boldsymbol{\varPi}$. The waveform
peak of $\hat{\ell}\left(\hat{\theta},\hat{\delta},p,d\right)$ in
Eq. \eqref{loglik1} locates at the true period $\hat{T}=\hat{p}/\left(df_{s}\right)$
and its multiples; see Fig. \ref{fig:likelihood} for an example.
Due to the existence of numerous local optimums, enumerating $p$
in $\boldsymbol{\mathcal{T}}$ is needed for optimization, i.e., 
\begin{equation}
\hat{p}=\arg\max_{p\in\boldsymbol{\mathcal{T}}}\left\{ \hat{\ell}\left(\hat{\theta},\hat{\delta},p,d\right)\right\} ,\label{eq:OPT}
\end{equation}
where $\boldsymbol{\mathcal{T}}=\left\{ 1,2,\cdots,dp_{max}\right\} $
is the searching range for $p$. Specifically, $\boldsymbol{\mathcal{I}}=\boldsymbol{\mathcal{T}}$
if $d=d_{*}$. Please refer to Supplementary Materials for a detailed
description and the pseudo codes for the proposed optimization of
Eqs. \eqref{eq:OPT0} and \eqref{eq:OPT}. The enumerating step for
$p$ can be computed in a parallel or distributed manner to further
reduce the computing time if applicable.

The settings of $\boldsymbol{\mathcal{I}}$ and $\boldsymbol{\mathcal{T}}$
depend on a balance between the desired estimation accuracy and the
available computational budget. When the sampling frequency $f_{s}$
is considerably high (that is, the number of points in one period
is reasonably large, e.g., larger than 100), setting $d=d_{*}=1$
(i.e., $\boldsymbol{\mathcal{T}}=\boldsymbol{\mathcal{I}}$) suffices,
which will lead to integer period estimates. Otherwise, we may set
$d>1$ and $d\geq d_{*}\geq1$ to enable period estimates having decimal
points. Empirical studies show that $d=d_{*}$ is not always necessary,
and $d_{*}=1$ suffices in all of our numerical studies. As the searching
range of $p$ will generally include more elements when dealing with
decimal periods than integer periods, computing decimal periods would
generally need more time, and thus some common choices suggested are
$d_{*}=1,d=5$ and $d_{*}=1,d=10$. The tuning parameter $p_{max}$
in $\boldsymbol{\mathcal{I}}$ and $\boldsymbol{\mathcal{T}}$ is
often chosen according to the prior knowledge of the signal, such
that $p_{max}/f_{s}$ is at least greater than the true period. For
example, as the fundamental frequency of voiced speeches ranges approximately
from 80 Hz to 300 Hz, we should set $p_{max}$ such that $f_{s}/p_{max}$
is smaller than 80 Hz. In addition, we could guess the true period
$T$ (and hence $p_{max}$) according to the waveform of the likelihood
function with a gradual increase of $p_{max}$, until the likelihood
exhibits approximately periodic local optimums.

It is worth to remark that computing the likelihoods of CPGP requires
a computational complexity of $\mathcal{O}\left(p^{2}\right)$ in
general and $\mathcal{O}\left(p\log p\right)$ when $p$ divides $n$,
while the total computational complexity of CPGP considering the estimation
of period is $\mathcal{O}\left(d^{3}p_{max}^{3}\right)$. As both
$p$ and $p_{max}$ are independent of and much smaller than $n$
in practice, CPGP remains scalable. In this work, we emphasize more
on the complexity for calculating likelihoods, because we want to
make a fair comparison between CPGP and current PGP-based models that
assume known periods. Although the proposed period estimation in CPGP
can be embedded into PGP, it is not applicable in practice, because
the optimization in \eqref{eq:OPT} often requires evaluating the
likelihood function thousands of times where both the PGP and TPGP
would be prohibitively slow.

\subsection{Scalable Model Prediction}

As shown by \citet{jones1998efficient}, the BLUP of $y\left(t\right)$
can be obtained by maximizing the joint likelihood $y\left(t\right)$
and $\boldsymbol{y}$ with regard to $y\left(t\right)$, that is,
$\hat{y}\left(t\right)=\arg\max_{y\left(t\right)}\mathsf{Pr}\left(y\left(t\right),\boldsymbol{y}\right)$.
Similar to the likelihood decomposition in Section~\ref{sec:ld},
the joint distribution $\mathsf{Pr}\left(y\left(t\right),\boldsymbol{y}\right)$
can be decomposed into two parts 
\[
\mathsf{Pr}\left(y\left(t\right),\boldsymbol{y}\right)=\mathsf{Pr}\left(y\left(t\right),\boldsymbol{y}_{*}|\boldsymbol{\varUpsilon}\right)\mathsf{Pr}\left(\boldsymbol{\varUpsilon}\right),
\]
where $\mathsf{Pr}\left(y\left(t\right),\boldsymbol{y}_{*}|\boldsymbol{\varUpsilon}\right)$
is the joint distribution of $\left\{ y\left(t\right),\boldsymbol{y}_{*}\right\} $
conditional on $\boldsymbol{\varUpsilon}$. Thus, we have 
\[
\hat{y}\left(t\right)=\arg\max_{y\left(t\right)}\mathsf{Pr}\left(y\left(t\right),\boldsymbol{y}_{*}|\boldsymbol{\varUpsilon}\right),
\]
and $\mathsf{Pr}\left(y\left(t\right),\boldsymbol{y}_{*}|\boldsymbol{\varUpsilon}\right)$
can be further written as 
\[
\left[\begin{array}{c}
y\left(t\right)\\
\boldsymbol{y}_{*}
\end{array}\right]\sim\left(\left[\begin{array}{c}
\mu\left(t\right)\\
\boldsymbol{\mu}
\end{array}\right],\sigma^{2}\left[\begin{array}{cc}
\pi\left(t\right) & \boldsymbol{\gamma}_{\bullet}^{T}\left(t\right)\\
\boldsymbol{\gamma}_{\bullet}\left(t\right) & \boldsymbol{\varPi}
\end{array}\right]\right),
\]
where 
\begin{equation}
\begin{cases}
\mu\left(t\right) & =\boldsymbol{f}^{T}\left(t\right)\boldsymbol{\beta}+\boldsymbol{\gamma}^{T}\left(t\right)\boldsymbol{\Sigma}^{-1}\left(\boldsymbol{\varUpsilon}-\boldsymbol{\varGamma}\right)\\
\pi\left(t\right) & =1+\delta^{2}-\boldsymbol{\gamma}^{T}\left(t\right)\boldsymbol{\Sigma}^{-1}\boldsymbol{\gamma}\left(t\right)\\
\boldsymbol{\gamma}_{\bullet}\left(t\right) & =\boldsymbol{\gamma}_{*}\left(t\right)-\boldsymbol{\Xi}^{T}\boldsymbol{\Sigma}^{-1}\boldsymbol{\gamma}\left(t\right)
\end{cases}.\label{eq:Cond}
\end{equation}
Here, $\boldsymbol{\gamma}\left(t\right)$ and $\boldsymbol{\gamma}_{*}\left(t\right)$
represent the correlation between $y\left(t\right)$ and $\boldsymbol{\varUpsilon}$
and the correlation between $y\left(t\right)$ and $\boldsymbol{y}_{*}$,
respectively. Therefore, given $\boldsymbol{\beta}$, $\sigma$, $\theta$,
$\delta,$ and $p$, we have the prediction $\hat{y}\left(t\right)=\mu\left(t\right)+\boldsymbol{\gamma}_{\bullet}^{T}\left(t\right)\boldsymbol{\varPi}^{-1}\left(\boldsymbol{y}_{*}-\boldsymbol{\mu}\right)$
and its variance $\mathrm{Var}(\hat{y}\left(t\right))=\sigma^{2}\left(\pi\left(t\right)-\boldsymbol{\gamma}_{\bullet}^{T}\left(t\right)\boldsymbol{\varPi}^{-1}\boldsymbol{\gamma}_{\bullet}\left(t\right)\right)$.

With Eq. \eqref{eq:Inv} plugged in, Eq. \eqref{eq:Cond} can be simplified
as 
\[
\begin{cases}
\mu\left(t\right) & =\boldsymbol{f}^{T}\left(t\right)\boldsymbol{\beta}+\frac{k}{\delta^{2}}\bar{\boldsymbol{\gamma}}^{T}\left(t\right)\boldsymbol{R}_{\delta}^{-1}\left(\bar{\boldsymbol{y}}-\bar{\boldsymbol{\varGamma}}\boldsymbol{\beta}\right)\\
\pi\left(t\right) & =1+\delta^{2}-\frac{k}{\delta^{2}}\bar{\boldsymbol{\gamma}}^{T}\left(t\right)\boldsymbol{R}_{\delta}^{-1}\bar{\boldsymbol{\gamma}}\left(t\right)\\
\boldsymbol{\gamma}_{\bullet}\left(t\right) & =\boldsymbol{\gamma}_{*}\left(t\right)-\frac{k}{\delta^{2}}\boldsymbol{R}_{\bullet}^{T}\boldsymbol{R}_{\delta}^{-1}\bar{\boldsymbol{\gamma}}\left(t\right)
\end{cases},
\]
where $\bar{\boldsymbol{\gamma}}\left(t\right)$ denotes the correlation
between $y\left(t\right)$ and $\boldsymbol{y}_{1}$. Then, by substituting
$\boldsymbol{\beta}$ with $\hat{\boldsymbol{\beta}}$, we can obtain
a computationally efficient formula of BLUP 
\begin{align}
\hat{y}\left(t\right) & =\boldsymbol{f}^{T}\left(t\right)\hat{\boldsymbol{\beta}}+\frac{k}{\delta^{2}}\bar{\boldsymbol{\gamma}}^{T}\left(t\right)\boldsymbol{R}_{\delta}^{-1}\left(\bar{\boldsymbol{y}}-\bar{\boldsymbol{\varGamma}}\hat{\boldsymbol{\beta}}\right)\label{eq:BLUP1}\\
 & +\boldsymbol{\gamma}_{\bullet}^{T}\left(t\right)\boldsymbol{\varPi}^{-1}\left(\boldsymbol{y}_{\bullet}-\boldsymbol{\varGamma}_{\bullet}\hat{\boldsymbol{\beta}}\right).\nonumber 
\end{align}

Eq. \eqref{eq:beta_hat} implies that $\mathrm{Var}(\hat{\boldsymbol{\beta}})=\sigma^{2}\left(\boldsymbol{S}_{\boldsymbol{\varGamma}\boldsymbol{\varGamma}}+\eta\boldsymbol{\varGamma}_{\bullet}^{T}\boldsymbol{\varPi}^{-1}\boldsymbol{\varGamma}_{\bullet}\right)^{-1}$,
and thus the variance of $\hat{y}\left(t\right)$, taking $\hat{\boldsymbol{\beta}}$
into account, is 
\begin{align*}
\mathrm{Var}\left(\hat{y}\left(t\right)\right) & =\sigma^{2}\left(\pi\left(t\right)-\boldsymbol{\gamma}_{\bullet}^{T}\left(t\right)\boldsymbol{\varPi}^{-1}\boldsymbol{\gamma}_{\bullet}\left(t\right)\right)\\
 & +\sigma^{2}\boldsymbol{\omega}^{T}\left(\boldsymbol{S}_{\boldsymbol{\varGamma}\boldsymbol{\varGamma}}+\eta\boldsymbol{\varGamma}_{\bullet}^{T}\boldsymbol{\varPi}^{-1}\boldsymbol{\varGamma}_{\bullet}\right)^{-1}\boldsymbol{\omega},
\end{align*}
where $\boldsymbol{\omega}=\boldsymbol{f}\left(t\right)-\frac{k}{\delta^{2}}\bar{\boldsymbol{\varGamma}}^{T}\boldsymbol{R}_{\delta}^{-1}\bar{\boldsymbol{\gamma}}\left(t\right)-\boldsymbol{\varGamma}_{\bullet}^{T}\boldsymbol{\varPi}^{-1}\boldsymbol{\gamma}_{\bullet}\left(t\right)$.

Similar to the above, the MATLAB toolbox ``smt\textquotedbl{} for
circulant matrices \citep{redivo2012smt} and the Cholesky factorization
algorithm for semidefinite Toeplitz matrices \citep{stewart1997cholesky}
can be applied to calculate this BLUP, which has a computational complexity
of $\mathcal{O}\left(p^{2}\right)$.

\section{Simulation Study \label{sec:Simulation}}

The PGP framework is mostly used in applications for periodicity detection.
Here, we conduct a popular simulation for periodicity detection to
evaluate the effectiveness of the proposed CPGP. We compare it with
some state-of-the-art methods, including TPGP \citep{zhang2005time},
FNLS \citep{nielsen2017fast}, NRC \citep{li2021extended}, and MLPE
\citep{wise1976maximum}. For a fair comparison, the proposed optimization
provided in Eqs. \eqref{eq:OPT0} and \eqref{eq:OPT} is also applied
in TPGP to enable period estimation. 

In addition, we also consider a variant of CPGP here, called the approximate
CPGP (ACPGP), in order to study the importance of $\ell_{2}$ in \eqref{eq:log-lik}.
The likelihood of ACPGP only includes the normalized $\ell_{1}$,
i.e., 
\begin{equation}
\ell\left(\theta,\delta,p,d\right)=\frac{1}{kp}\ell_{1}.\label{eq:ACPGP}
\end{equation}
In ACPGP, the MLE of $\boldsymbol{\beta}$ and $\sigma^{2}$ is $\hat{\boldsymbol{\beta}}=\boldsymbol{S}_{\boldsymbol{\varGamma}\boldsymbol{\varGamma}}^{-1}\boldsymbol{S}_{\boldsymbol{\varGamma}\boldsymbol{Y}}$
and $\hat{\sigma}^{2}=\frac{1}{n}\left(\boldsymbol{S}_{\boldsymbol{Y}\boldsymbol{Y}}-\hat{\boldsymbol{\beta}}^{T}\boldsymbol{S}_{\boldsymbol{\varGamma}\boldsymbol{\varGamma}}\hat{\boldsymbol{\beta}}\right)$,
respectively. Similar to Eq. \eqref{loglik1}, we have $\hat{\ell}\left(\hat{\theta},\hat{\delta},p,d\right)=-\frac{1}{2kp}\left(\log\hat{\sigma}^{2kp}\hat{\delta}^{2kp}+\log\left|\hat{\boldsymbol{R}}_{\hat{\delta}}\right|+kp\right),$
and the same optimization algorithm in Eqs. \eqref{eq:OPT0} and \eqref{eq:OPT}
can be used for parameter estimation. Notably, computing the likelihood
of ACPGP requires $\mathcal{O}\left(p\log p\right)$ time, which is
generally faster than CPGP at the price of sacrificing certain accuracy.

Consider a widely discussed simulation on the synthetic signals in
\citet{fan2018noise}, which are the periodic transients generated
from the following equation: 
\begin{equation}
x\left(t\right)=\sum_{i=0}^{\left\lfloor \frac{l}{T_{0}}\right\rfloor }e^{\frac{-\zeta2\pi\omega\left(t-iT_{0}\right)^{2}}{\sqrt{1-\zeta^{2}}}}\sin2\pi\omega\left(t-iT_{0}\right),\label{eq:syn_signal}
\end{equation}
where $\zeta=0.01$ is the damping ratio, $\omega=0.055$ Hz is the
natural frequency, the period is $T_{0}=200$ seconds, and $l$ is
the time length of the signals. The signals $x(t)$ are collected
with a sampling frequency of $f_{s}=1$ Hz, and thus there are 200
signal points in one period. The contaminated signal $y(t)$ (i.e.,
the simulated data) is generated by adding Gaussian white noises to
the periodic signal $x(t)$. The focus here is periodicity detection.
%We compare the performance of CPGP for periodicity detection with those of some state-of-the-art methods, including TPGP, the FNLS estimator \citep{nielsen2017fast}, the NRC method \citep{li2021extended}, and the MLPE method \citep{wise1976maximum}.

In CPGP and ACPGP, as there are 200 points in one period, the searching
range of $p$ is set to $\left[1,500\right]$, that is, $\boldsymbol{\mathcal{I}}=\boldsymbol{\mathcal{T}}=\left\{ 1,2,\cdots,500\right\} $,
$p_{max}=500$ and $d_{*}=d=1$. The searching range for the parameters
$\delta$ and $\theta$ is set to $\left[2,20\right]$ and $\left[1,30\right]$,
respectively. Their initial values in the optimization algorithm are
assigned to those that have the maximum value of $\hat{\ell}_{\boldsymbol{\mathcal{I}}}\left(\theta,\delta,d_{*}\right)$
among nine grid candidates in the two-dimensional searching space
$\left[2,20\right]\times\left[1,30\right]$. Using the best initial
values from a simple grid search would make CPGP more robust, especially
when prior information (domain knowledge) is unavailable, as the likelihood
may have many misleading local optimums. Throughout this paper, we
use $\boldsymbol{f}(t)=1$ in CPGP for simplicity. In this study,
both CPGP and TPGP use the same initial parameter settings, and thus
they result in the same parameter estimations. Their difference lies
in the computing time, where CPGP is much faster than TPGP. In FNLS,
the max harmonic number is set to $L=30$. We replicate the analysis
100 times where the signals are independently and randomly generated.

\begin{figure}[t]
\begin{centering}
\includegraphics[width=9cm]{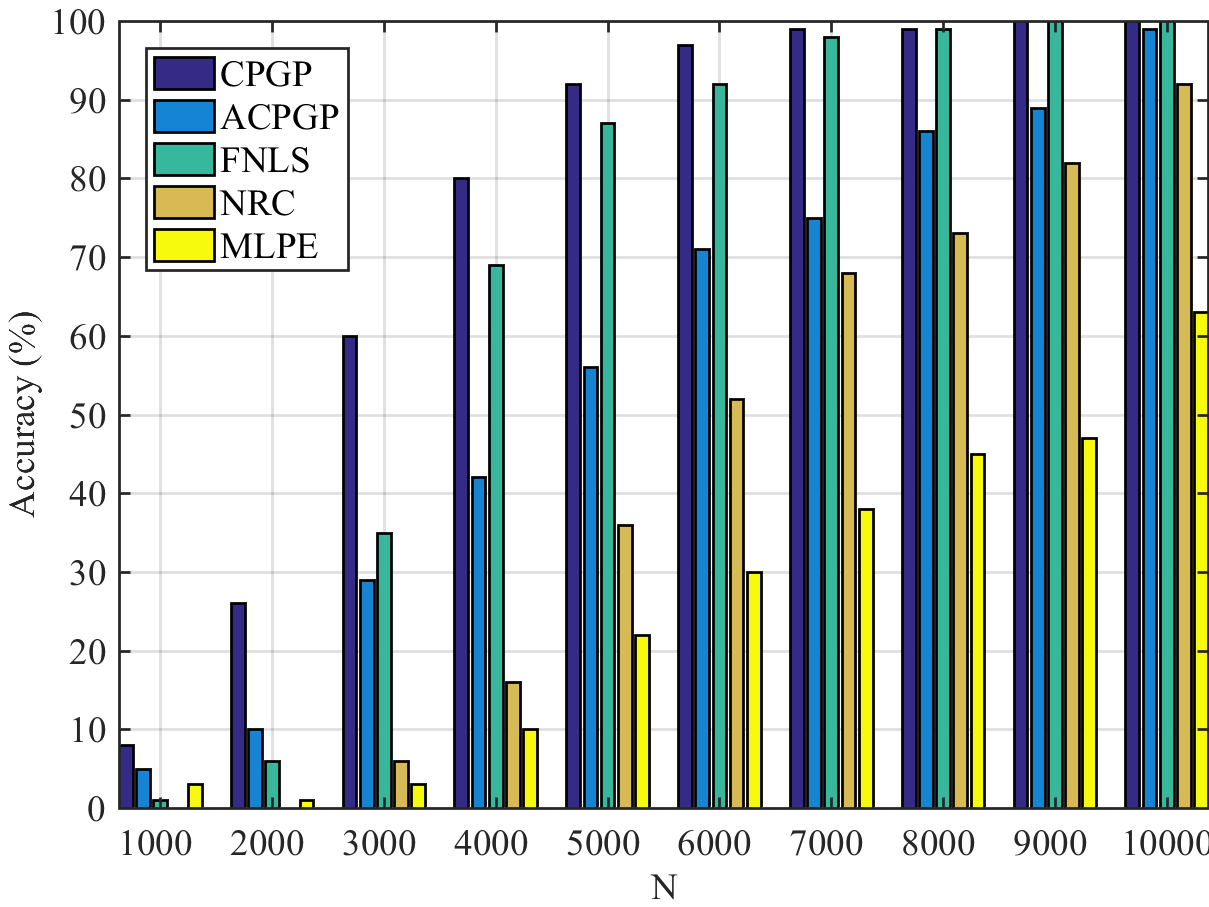} 
\par\end{centering}
\caption{\label{fig:simulation_len}Period estimation accuracy of CPGP, {ACPGP,}
FNLS, NRC, and MLPE changing with the signal length in the simulation
study.}
\end{figure}

\begin{figure}[t]
\begin{centering}
\includegraphics[width=9cm]{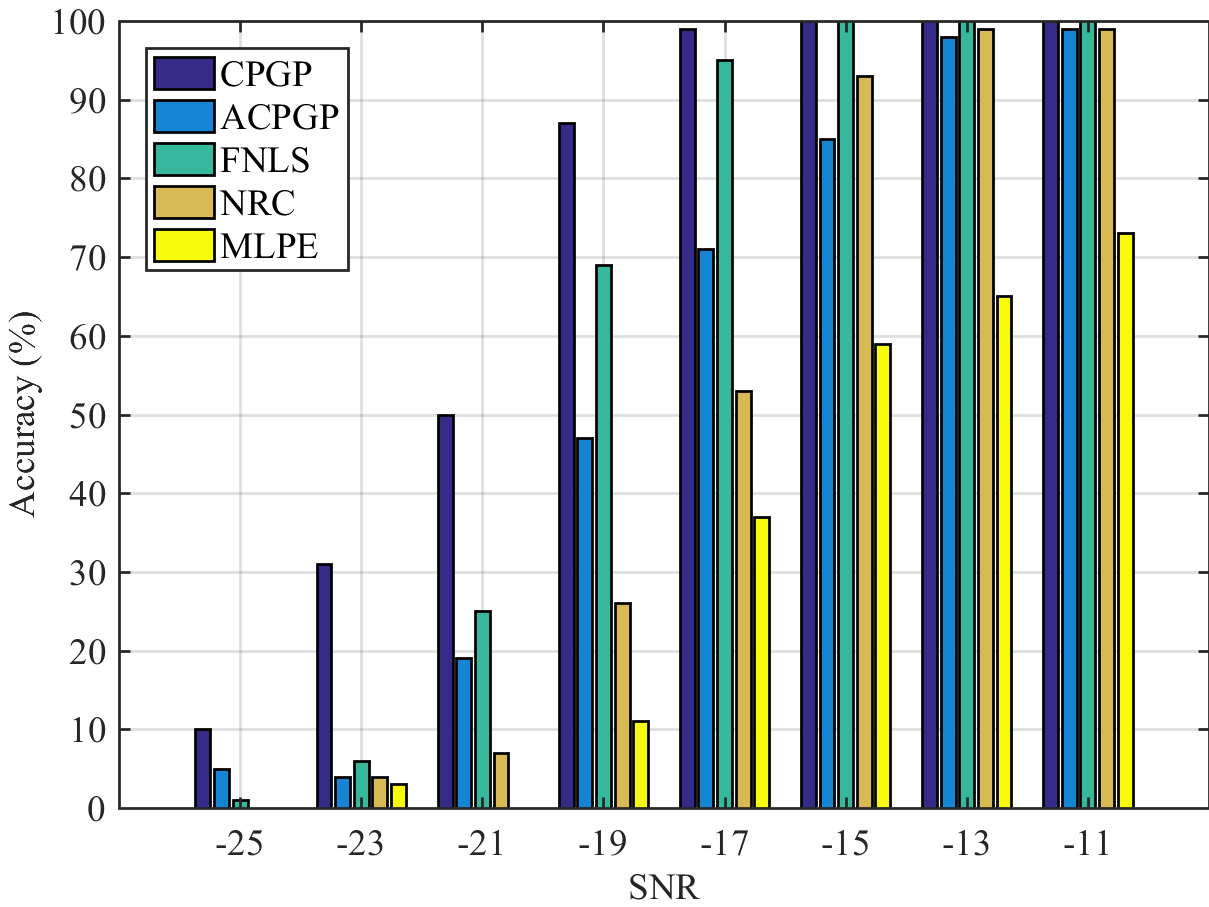} 
\par\end{centering}
\caption{\label{fig:simulation_SNR}Period estimation accuracy of CPGP, ACPGP,
FNLS, NRC, and MLPE changing with SNR in the simulation study.}
\end{figure}

In Fig.~\ref{fig:simulation_len}, we show the period estimation
accuracy (i.e., the probability of $\hat{T}=T_{0}$ out of the 100
trials) of each method handling various signal lengths (from 1000
to 10000), where the SNR is $-21$ dB. In Fig.~\ref{fig:simulation_SNR},
we show the period estimation accuracy of each method handling signals
with various SNRs (from $-25$ dB to $-11$ dB), where the signal
length is 5000. Notably, as TPGP has the same period estimates as
CPGP, we do not show its results in these figures. Their difference
lies in the computing time which is shown in Tab.~\ref{tab:ComputationTime}.

In Figs.~\ref{fig:simulation_len} and \ref{fig:simulation_SNR},
compared with CPGP, ACPGP reports a lower period estimation accuracy
in all cases, which suggests that the second composite likelihood
$\ell_{2}$ in \eqref{eq:log-lik} may play an important role here.
It is seen that ACPGP only works reasonably well for long signals
and increased SNR, but works badly for other cases. In practice, we
only promote the use of CPGP for period detection due to its high
accuracy and robustness. If we know in advance that the signal length
will be very long and SNR will be high enough, ACPGP can be used considering
its faster speed.

Additionally, it is seen from Fig. \ref{fig:simulation_len} that
CPGP has substantially higher accuracy than FNLS, NRC, and MLPE for
all signal lengths. As the signal length increases, all methods would
eventually reach 100\% accuracy, while CPGP converges considerably
faster. Clearly, CPGP outperforms its competitors and is able to accurately
detect periodicity at an early stage. In addition, from Fig. \ref{fig:simulation_SNR},
we can see that CPGP has substantially higher accuracy than other
methods when the SNR is low, and all methods would have higher accuracy
as the SNR increases. Above all, CPGP can process lower SNR signals
more effectively via using fewer signal points compared with FNLS,
NRC, and MLPE.

\begin{figure}[t]
\begin{centering}
\includegraphics[width=9.5cm]{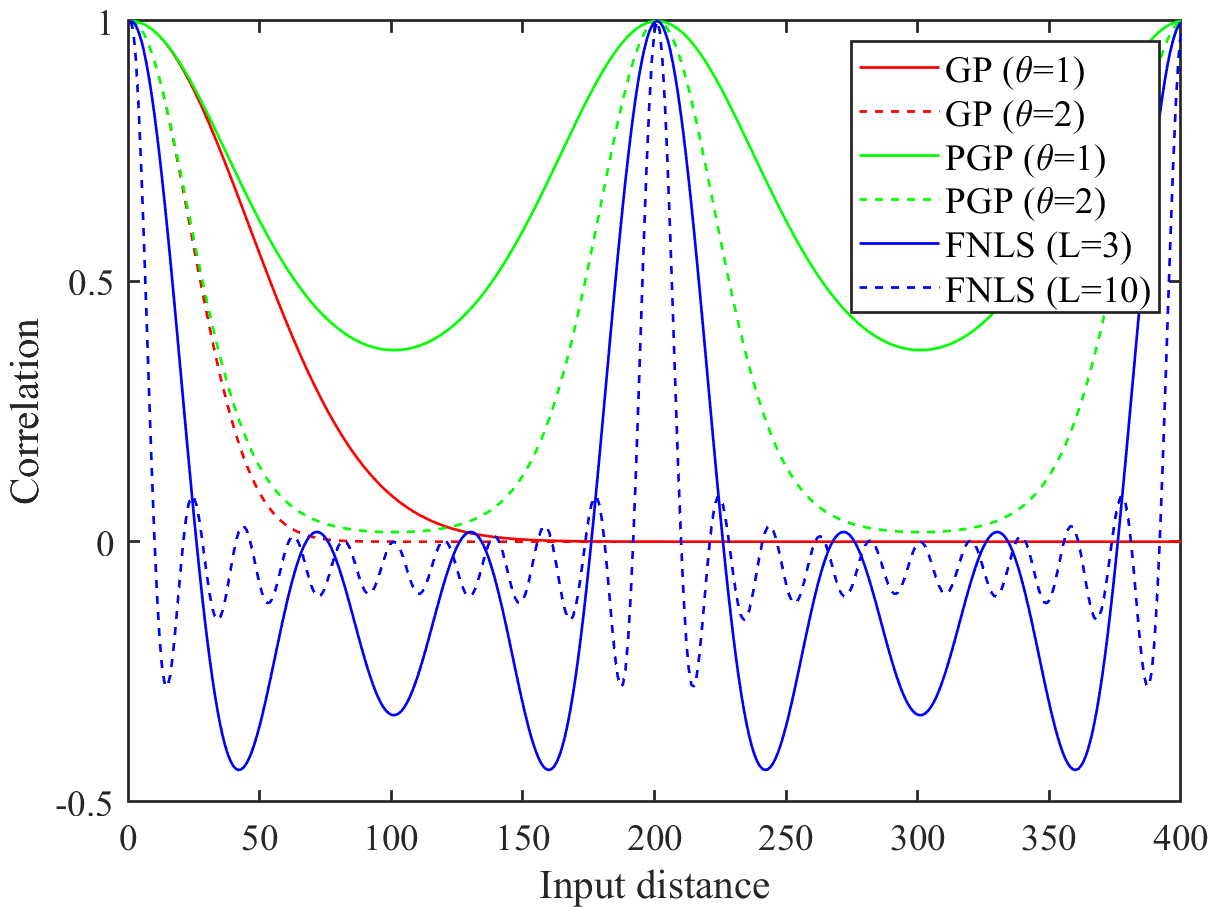} 
\par\end{centering}
\caption{\label{fig:correlation}Correlation function of GP, PGP (CPGP), and
FNLS}
\end{figure}

Notably, NRC and MLPE provide lower accuracy mainly because they do
not consider the within-period correlations, which are meticulously
modeled by CPGP. Different from FNLS that uses a number of Fourier
bases to approximate the true signals, CPGP uses a non-parametric
way to represent the signals where the roughness parameter $\theta$
is optimized to adaptively model the within-period correlations. As
shown in Fig. \ref{fig:correlation}, the correlation function of
PGP (and CPGP) is more flexible than that of FNLS. Most correlations
in FNLS are quite small, and diagonal elements dominate the off-diagonal
ones in the correlation matrix. Most of these off-diagonal correlation
elements are negative, especially when $L$ is large. Yet, it should
be more reasonable and flexible to assume a zero (or near zero) correlation
for two distant points. In addition, the position of these negative
elements in the covariance matrix of FNLS is fixed given the tuning
parameter $L$ (although $L$ can be optimized via model selection
criteria). Clearly, such a correlation structure in FNLS may not effectively
utilize the spatial or temporal correlations of periodic data. The
superior performance of CPGP is mainly due to its capability to model
the circulant within-period correlations of periodic data. 

In Tab. \ref{tab:ComputationTime}, we report the average computing
time of 100 trials for each method dealing with various signal lengths.
Here, TPGP is the slowest, and it may require several hours for modeling
signals with only moderate lengths (e.g. $n=5000$). Notably, TPGP
is much faster than the classic PGP model, and similarly fast compared
to other PGP-based methods using approximation approaches discussed
in Section~\ref{sec:review}. In contrast, the proposed CPGP is significantly
faster than TPGP, which only requires around 18 seconds even for very
long signals. As seen in Tab.~\ref{tab:ComputationTime}, the computing
time of CPGP and ACPGP does not increase along with the signal length,
which indicates that the computational complexities of CPGP and ACPGP
are scalable. ACPGP is faster but less accurate compared to CPGP.

In Tab. \ref{tab:ComputationTime}, although FNLS consumes less time
than CPGP for short signals, it requires more time for long signals
(e.g., $n\geq10,000$). This is because its computational complexity
is $\mathcal{O}\left(nL\right)$ which grows with the signal length.
Compared to CPGP and FNLS, NRC and MLPE are faster but much less accurate
(as shown in Figs.~\ref{fig:simulation_len} and \ref{fig:simulation_SNR}),
since they ignore within-period correlations.

\begin{table}[t]
\centering{}\caption{\label{tab:ComputationTime}Average computing time of 100 trials (unit:
second).}
\begin{tabular}{ccccccc}
\hline 
{\small{}$n$}  & {\small{}TPGP }  & {\small{}CPGP}  & {\small{}ACPGP}  & {\small{}FNLS }  & {\small{}NRC }  & {\small{}MLPE}\tabularnewline
\hline 
{\small{}1000 }  & {\small{}189.91 }  & {\small{}18.744 }  & {\small{}5.9217}  & {\small{}0.6314 }  & {\small{}0.0117 }  & {\small{}0.0198}\tabularnewline
{\small{}2000 }  & {\small{}897.01}  & {\small{}18.960}  & {\small{}5.6760}  & {\small{}1.6991 }  & {\small{}0.0113 }  & {\small{}0.0216}\tabularnewline
{\small{}3000 }  & {\small{}1771.5 }  & {\small{}18.809 }  & {\small{}5.7584}  & {\small{}3.6112 }  & {\small{}0.0139 }  & {\small{}0.0247}\tabularnewline
{\small{}4000 }  & {\small{}3167.2 }  & {\small{}19.308 }  & {\small{}6.0763}  & {\small{}5.6430 }  & {\small{}0.0177 }  & {\small{}0.0304}\tabularnewline
{\small{}5000 }  & {\small{}4899.6 }  & {\small{}19.634 }  & {\small{}5.9567}  & {\small{}7.6847 }  & {\small{}0.0199 }  & {\small{}0.0328}\tabularnewline
{\small{}6000 }  & {\small{}6129.8 }  & {\small{}18.307 }  & {\small{}5.9955}  & {\small{}10.418 }  & {\small{}0.0216 }  & {\small{}0.0343}\tabularnewline
{\small{}7000 }  & {\small{}8324.0 }  & {\small{}18.056 }  & {\small{}5.9258}  & {\small{}11.620 }  & {\small{}0.0239 }  & {\small{}0.0359}\tabularnewline
{\small{}8000 }  & {\small{}10878 }  & {\small{}18.172 }  & {\small{}6.0060}  & {\small{}14.971 }  & {\small{}0.0265 }  & {\small{}0.0399}\tabularnewline
{\small{}9000 }  & {\small{}13848 }  & {\small{}17.935 }  & {\small{}6.0506}  & {\small{}15.912 }  & {\small{}0.0313 }  & {\small{}0.0453}\tabularnewline
{\small{}10000 }  & {\small{}17076}  & {\small{}18.337 }  & {\small{}6.1132}  & {\small{}20.098 }  & {\small{}0.0341 }  & {\small{}0.0488}\tabularnewline
\hline 
\end{tabular}
\end{table}

\section{Real Case Study \label{sec:Case-Study}}

\subsection{Bearing Fault Detection}

As one of the core components in rotating machinery, the rolling bearing
directly affects the product quality and reliability of the rotating
machines. However, bearings are prone to failure under complex working
conditions, thus triggering periodic vibrations. A common solution
to bearing fault detection is to apply period estimation methods to
extract the fault period from the vibration signals. Accurate methods
are needed to identify the fault period as early as possible because
the complicated background noise may mask the periodic signals of
the faulty bearing.

In this case study, the vibration data are downloaded from the Case
Western Reserve University (CWRU) \footnote{https://engineering.case.edu/bearingdatacenter/}.
The type of bearings used is 6205-2RS JEM SKF, and the parameters
of the test bearing are listed in Tab. \ref{tab:PARAMETERS-OF-TEST}.
The bearing fault is set in the outer race, and the signals are collected
from the testing rig with a sampling frequency of 48 kHz. According
to Tab. \ref{tab:PARAMETERS-OF-TEST}, the fault characteristic frequency
is 104 Hz, and the fault period is $T_{0}=461$. 
\begin{table}[t]
\caption{\label{tab:PARAMETERS-OF-TEST}Parameters of test bearing 6205-2RS
JEM SKF}

\centering{}%
\begin{tabular}{|c|c|}
\hline 
Type  & 6205-2RS JEM SKF\tabularnewline
\hline 
No. of rolling elements  & 9\tabularnewline
\hline 
Inside diameter (in)  & 0.9843\tabularnewline
\hline 
Outside diameter (in)  & 2.0472\tabularnewline
\hline 
Thickness (in)  & 0.5906\tabularnewline
\hline 
Ball diameter (in)  & 0.3126\tabularnewline
\hline 
Pitch diameter (in)  & 1.537\tabularnewline
\hline 
Approx. motor speed (rpm)  & 1747\tabularnewline
\hline 
Size of fault (mm)  & 0.021\tabularnewline
\hline 
Approx. fault frequency (Hz)  & 104\tabularnewline
\hline 
Approx. fault period (ms)  & 9.6\tabularnewline
\hline 
\end{tabular}
\end{table}

\begin{figure}[t]
\begin{centering}
\includegraphics[width=9.5cm]{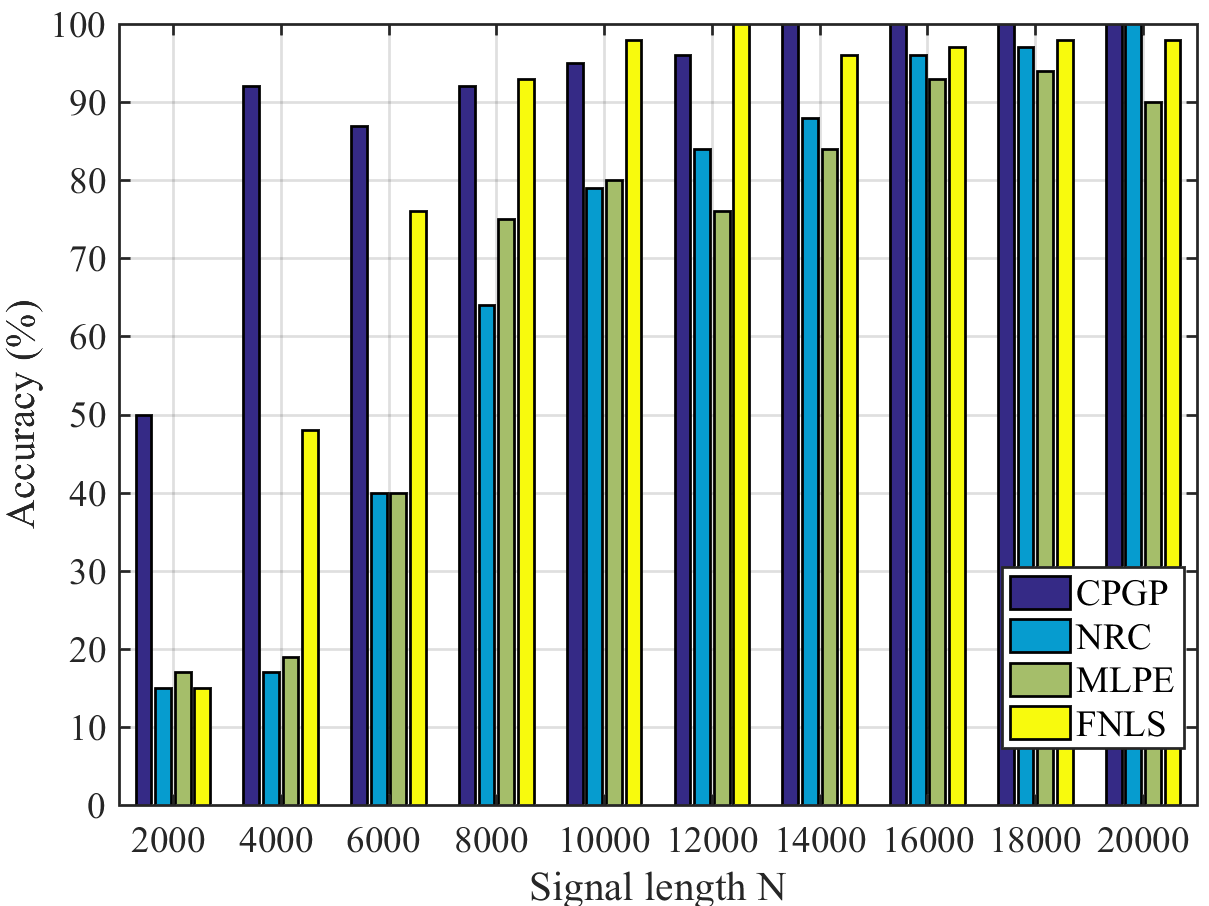} 
\par\end{centering}
\caption{\label{fig:CWRU_accuracy}Period estimation accuracy of CPGP, NRC,
MLPE, and FNLS on bearing vibration signals.}
\end{figure}

The vibration signals have 100,000 sample points. The fault characteristic
frequency cannot be directly located through conventional spectrum
methods due to the complex background noise. Thus, we seek to detect
the bearing fault from the time domain. The proposed CPGP method is
applied to estimate the fault period from the vibration signals. As
the sampling frequency is considerably high, the searching range for
$p$ is set to $\left[1,500\right]$, that is, $\boldsymbol{\mathcal{T}}=\boldsymbol{\mathcal{I}}=\left\{ 1,2,\cdots,1,000\right\} $,
$p_{max}=1,000$ and $d=d_{*}=1$. The searching range for parameters
$\delta$ and $\theta$ in CPGP is set to $\left[0.01,5\right]$ and
$\left[15,20\right]$, respectively. The initial values of $\delta$
and $\theta$ are obtained in a similar way to that in Section~\ref{sec:Simulation}.
To perform a comparison, NRC, MLPE, and FNLS are applied to estimate
the fault period. The max harmonic number for FNLS is set to $L=30$.
To measure the accuracy of each method, a segment of signals is randomly
drawn from the entire vibration signals with the signal length $n$
ranging from 2,000 to 20,000. Following \citet{li2021extended}, the
probability that the estimated period falls into $\left\{ 0.95T_{0},1.05T_{0}\right\} $
is recorded as the detection accuracy. The experiment is repeated
100 times for each $n$ ranging from 2,000 to 20,000, and we show
their fault detection accuracy in Fig. \ref{fig:CWRU_accuracy}. 

Fig. \ref{fig:CWRU_accuracy} shows that the proposed CPGP performs
substantially better than other methods, especially when the signal
length is short (smaller than 6,000), which indicates that it can
be used to detect the bearing fault at an early stage. Compared with
the accuracy of NRC and MLPE, the accuracy of CPGP is substantially
higher, and it converges to 100\% much faster as the signal length
increases (longer than 14,000). Although the FNLS slightly outperforms
CPGP when the signal length is moderate, its accuracy cannot converge
to 100\% even when the signals are sufficiently long. This is mainly
because the real vibration signals are not strictly periodic and subject
to significant noises. The steady convergence of CPGP indicates that
it can perform robustly on real signals.

\subsection{Pitch Estimation}

Pitch estimation, which is also known as fundamental frequency estimation,
is an intriguing research topic in the audio signal processing area
\citep{nielsen2017fast}. Researchers aim to produce a sequence of
frequency values corresponding to the pitch of speech or musical recordings.
Given a recording of monophonic music or speech, its pitch could be
easily identified by humans. However, completing such a task is still
challenging for AI because the signals usually comprise complex signal
components and environment noises. The pitch can be estimated by the
autocorrelation-based methods (e.g., NRC) or the likelihood-based
methods (e.g., MLPE). Yet, NRC and MLPE do not consider within-period
correlations. A more complex method that can take within-period correlations
into consideration is needed. In this case study, CPGP is applied
to estimate the pitch of a voice speech recording and a monophonic
music recording produced by the viola. 
\begin{figure}[t]
\begin{centering}
\includegraphics[width=9.5cm]{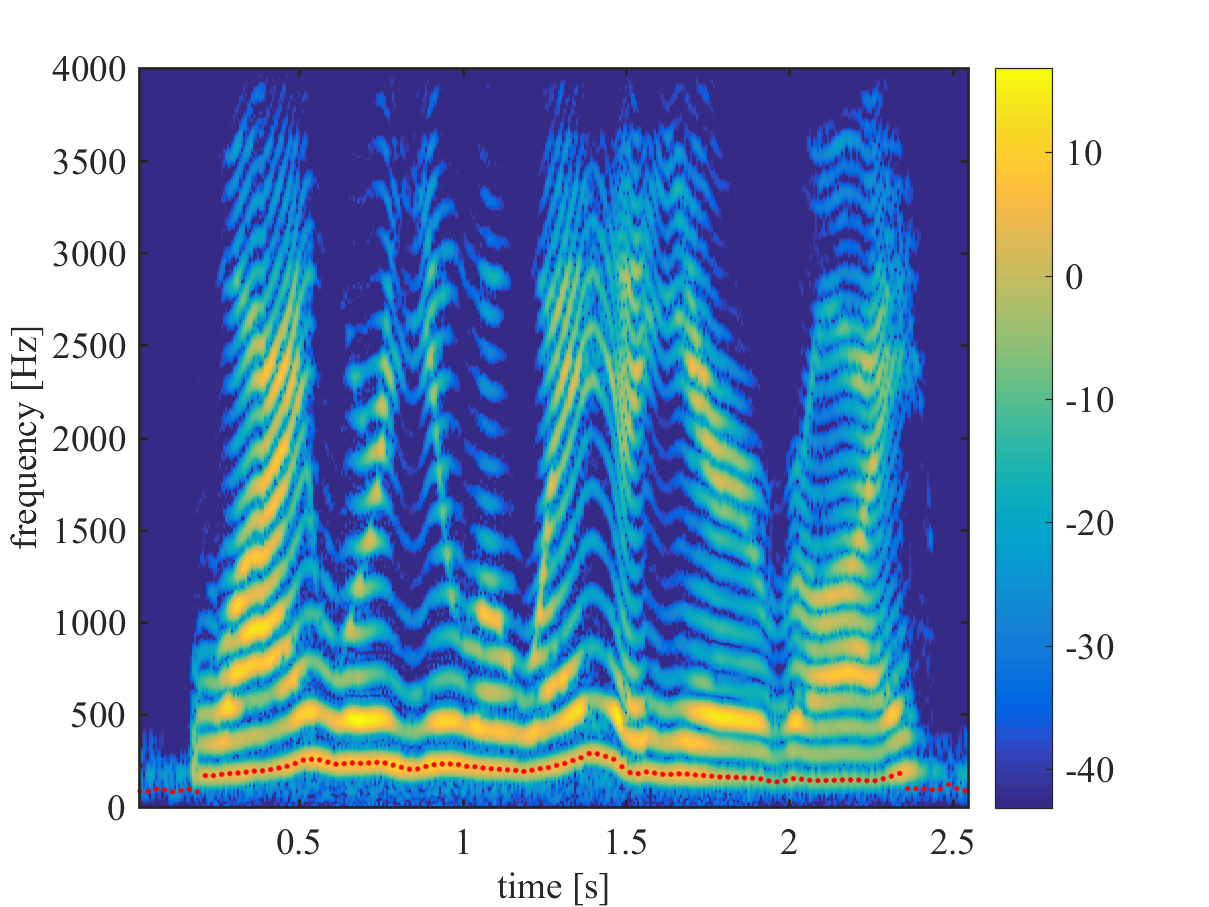} 
\par\end{centering}
\caption{\label{fig:SpeechSpec}Estimated fundamental frequencies and spectrograms
of the speech signals.}
\end{figure}

The signals are divided into short signal subsets ($n=200$ for the
speech signals and $n=240$ for the music signals), and each subset
is analyzed separately to display the flow of changing fundamental
frequencies. The searching grids for $p$ are $\boldsymbol{\mathcal{I}}=\left\{ 2,3,\cdots,120\right\} $
and $\boldsymbol{\mathcal{T}}=\left\{ 2,3,\cdots,600\right\} $ via
setting $d_{*}=1$, $d=5$ and $p_{max}=120$, because the sampling
frequencies (8000 Hz) of these signals are relatively low. As the
waveform of audio signals is considerably smoother and the strength
of noises is weak, the searching range for parameters $\delta$ and
$\theta$ in CPGP is set to $\left[0.3,0.5\right]$ and $\left[2,3\right]$,
respectively. We find that using the domain knowledge to specify searching
ranges would help improve the performance.

\begin{figure}[t]
\begin{centering}
\includegraphics[width=9.5cm]{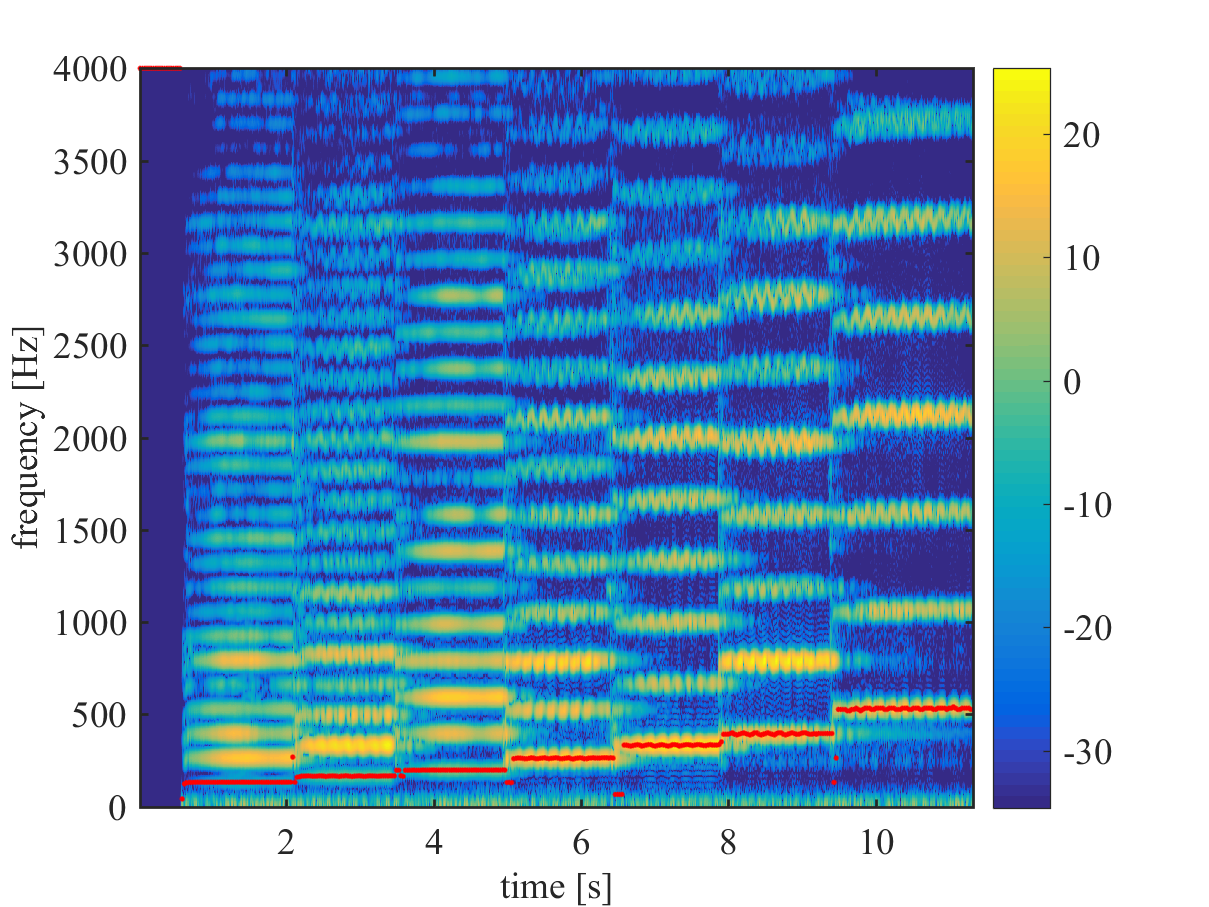} 
\par\end{centering}
\caption{\label{fig:MonoSpec}Estimated fundamental frequencies and spectrograms
of the musical signals.}
\end{figure}

As shown in Figs. \ref{fig:SpeechSpec} and \ref{fig:MonoSpec}, the
estimated fundamental frequencies for both signals are marked as red
spots on the spectrograms \citep{mitra2006digital}. The red spots
match well with the patterns illustrated in the spectrograms, which
indicates that CPGP can accurately estimate the period of audio and
monophonic musical signals. This case study implies the great potential
of applying CPGP in pitch estimation and melody extraction for audio
and musical signals. 

\section{Conclusion \label{sec:conc}}

In this paper, a scalable modeling approach called CPGP is proposed
to accelerate the parameter estimation and model prediction of PGP,
thereby greatly substantiating its applicability for handling large-scale
and low SNR periodic data. When periodic data are collected at grids,
computing the likelihoods of CPGP only requires a computational complexity
of $\mathcal{O}\left(p^{2}\right)$ in general, and $\mathcal{O}\left(p\log p\right)$
if $p$ divides $n$; its total computational complexity considering
the estimation of periods is $\mathcal{O}\left(d^{3}p_{max}^{3}\right)$,
where the segment lengths $p$ and $p_{max}$ are independent of and
much smaller than the data size $n$ in practice. Different from conventional
approximation approaches, CPGP has exactly the same likelihoods and
thus maintains the same accuracy as PGP. The superior performances
of CPGP in the simulations and case studies validates its usefulness
in real-world applications.

An interesting future topic is how to further accelerate CPGP with
desirable accuracy. In the simulation study, it is seen that the variant
ACPGP runs faster than CPGP, but its performance is not satisfactory.
Simply sacrificing the second composite log-likelihood $\ell_{2}$
to improve the speed is not a good idea. An alternative approach is
to apply zero-padding of the data such that $n$ can divide $p$,
which will accelerate the computational complexity of evaluating likelihoods
to $\mathcal{O}\left(p\log p\right)$. Yet, a straightforward application
of zero-padding does not perform well in practice, especially when
signals have non-constant trends. A meticulously designed zero-padding
method with solid theoretical supports will be explored in our future
works, which uses the circulant embedding method \citep{wood1994simulation,davies2013circulant}
and the circulant approximation for the Toeplitz matrix.

The proposed CPGP assumes the data are strictly periodic, although
it has been successfully applied to real applications in which signals
are probably pseudo-periodic. A new efficient modeling approach may
be required for pseudo-periodic signals with strong stretched or distorted
variations of a repeating cycle. Another restrictive assumption is
the noises are white Gaussian. These assumptions may limit the applications
of the proposed CPGP. We will explore some extensions of CPGP to pseudo-periodic
data contaminated with colored noises in future research.

\vspace{0.25in}

\begin{center}
\textbf{\Large{}Supplementary Materials}{\Large\par}
\par\end{center}

The supplementary materials contain technical proofs of the theoretical results, details (including pseudo codes) of the proposed optimization for parameter estimation, and codes for reproducing all figures and tables in this article. 
\vspace{0.25in}

\bibliographystyle{asa}
\bibliography{jos}

\end{document}